%% file: ms.edit.tex
\documentclass[traditabstract]{aa}

\usepackage{graphicx}
\usepackage{epstopdf}
\usepackage{deluxetable}
\usepackage{times,epsfig} 
\usepackage{mathrsfs}  
\usepackage{natbib}
\usepackage{amssymb}
\bibpunct{(}{)}{;}{a}{}{,} 
\usepackage{caption}

\usepackage[english]{babel}
\usepackage{longtable}
\usepackage{url}
\usepackage{hyperref}

\authorrunning{Stritzinger et al.}
\titlerunning{An observational study on the faint and fast SN~2010ae.}
\begin{document}

\title{Optical and near-IR observations of the faint and fast 2008ha-like supernova 2010ae\thanks{Based on observations
 collected at the European Organization for Astronomical Research in the Southern Hemisphere, Chile (ESO Program 082.A--0526, 084.D--0719,
 088.D-0222, 184.D-1140, and 386.D--0966;  
 the Gemini Observatory, Cerro Pachon, Chile (Gemini Program GS-2010A-Q-14 and GS-2010A-Q-38); the Magellan 6.5 m telescopes at  Las Campanas Observatory; and the SOAR telescope.}}

\author{M.~D. Stritzinger\inst{1}
\and E. Hsiao\inst{2}
\and S. Valenti\inst{3}
\and F. Taddia\inst{4}
\and T. J. Rivera-Thorsen\inst{4}
\and G. Leloudas\inst{5,6}
\and K. Maeda\inst{7,8}
\and A. Pastorello\inst{9}
\and M.~M. Phillips\inst{2}
\and G. Pignata\inst{10}
\and E. Baron\inst{11}
\and C.~R. Burns\inst{12}
\and C. Contreras\inst{1,2}
\and G. Folatelli\inst{7}
\and M. Hamuy\inst{13}
\and P. H\"{o}eflich\inst{14}
\and N. Morrell\inst{2}
\and J. L. Prieto\inst{15}
\and S. Benetti\inst{8}
\and A. Campillay\inst{2}
\and J. B. Haislip\inst{16}
\and A. P. LaClutze\inst{16}
\and J. P. Moore\inst{16}
\and D. E. Reichart\inst{16}
}

\institute{Department of Physics and Astronomy, Aarhus University, Ny Munkegade 120, DK-8000 Aarhus C, Denmark\\ (\email{max@phys.au.dk})
\and
Carnegie Observatories, Las Campanas Observatory, 
  Casilla 601, La Serena, Chile
\and
Las Cumbres Observatory Global Telescope Network, Inc.
Santa Barbara, CA 93117, USA
\and
The Oskar Klein Centre, Department of Astronomy, Stockholm University, AlbaNova, 10691 Stockholm, Sweden
\and
The Oskar Klein Centre, Department of Physics, Stockholm University, AlbaNova, 10691 Stockholm, Sweden
 \and
 Dark Cosmology Centre, Niels Bohr Institute, University of Copenhagen, Juliane Maries Vej 30, 2100 Copenhagen \O, Denmark
 \and
 Department of Astronomy, Kyoto University, Kitashirakawa-Oiwake-cho,
Sakyo-ku, Kyoto 606-8502, Japan
\and
 Kavli Institute for the Physics and Mathematics of the Universe (IPMU), University of Tokyo, 5-1-5 Kashiwanoha, Kashiwa, Chiba 277-8583, Japan
\and
INAF-Osservatorio Astronomico di Padova, vicolo
dell Osservatorio 5, 35122 Padova, Italy
\and
Departamento Ciencias Fõsicas, Universidad Andres Bello, Av. Republica 252, Santiago, Chile
\and
Homer L. Dodge Department of Physics and Astronomy, University of Oklahoma, 440 W. Brooks, Rm 100, Norman, OK 73019-2061, USA
\and
Observatories of the Carnegie Institution for Science, 813 Santa Barbara St., Pasadena, CA 91101, USA
\and
Departamento de Astronomia, Universidad de Chile, Casilla 36D, Santiago, Chile
\and
Department of Physics, Florida State University, Tallahassee, FL 32306, USA
\and
Department of Astrophysical Sciences, Princeton University, NJ 08544, USA
\and
University of North Carolina at Chapel Hill, Campus Box 3255, Chapel Hill, NC 27599-3255, USA
}

\date{Received 22 October 2013 / Accepted 18 November 2013}

\abstract{A comprehensive set of optical and near-infrared (NIR) photometry and spectroscopy  is presented for the faint and fast 2008ha-like supernova (SN)~2010ae.
Contingent on the adopted value of host extinction, SN~2010ae reached a peak brightness of $-13.8 > M_V > -15.3$ mag, while modeling of the UVOIR light curve  suggests it produced 
0.003--0.007~$M_{\sun}$ of $^{56}$Ni, ejected 0.30--0.60~$M_{\sun}$ of material, and had an explosion energy of 0.04--0.30$\times$10$^{51}$~erg.
The values of these explosion parameters are similar to  the peculiar SN~2008ha --for which we also present previously unpublished early 
phase optical and NIR light curves-- and places these two transients at the faint end of the 2002cx-like SN population. 
Detailed inspection of the post maximum NIR spectroscopic sequence indicates the presence of a multitude of spectral features, which are identified through {\tt SYNAPPS}  modeling to be mainly attributed to $\ion{Co}{ii}$.  Comparison with a collection of 
 published and unpublished NIR spectra of  other 2002cx-like SNe,
  reveals that a $\ion{Co}{ii}$ footprint is  ubiquitous  to this subclass of transients, 
  providing a link to Type~Ia SNe. 
  A visual-wavelength spectrum of SN~2010ae obtained at $+$252 days past maximum shows a striking resemblance to a similar epoch spectrum of SN~2002cx. 
However, subtle differences in the strength and ratio of calcium emission features, as well 
as diversity among similar epoch spectra of other 2002cx-like SNe indicates 
a range of physical  conditions of the  ejecta,  highlighting the heterogeneous nature of this peculiar class of transients.
}

\maketitle
\section{Introduction}
\label{sec:intro}

In recent years dedicated transient search programs have led to the tantalizing 
discovery of a  diverse population of hydrogen deficient, rapidly evolving supernovae (SNe). 
Detailed studies of  these  new transients  have lead to the emergence of  
a handful of sub-classes whose origins are a matter of open debate.
For instance, SNe~2002bj and 2010X had extremely short rise times
($\tau_{R} \leq$ 7 days), reached peak absolute $R$-band magnitudes 
($M_{R}$) of  $-18$ and  $-$17 mag, respectively, and exhibited
expansion velocities on the order of 10$^{4}$ km~s$^{-1}$; altogether 
suggestive  of progenitors with very small ejected masses, i.e., $<$ 0.5 M$_{\sun}$   \citep{poznanski10,kasliwal10,perets11}.
Another group of fast evolving transients  recently identified 
consists of,  amongst others, SN~2005E \citep{perets10}, PTF~09dav \citep{sullivan11,kasliwal12}, and SN~2012hn \citep{valenti13}. 
These objects share the characteristics of low peak luminosities ranging between $M_{R} \sim -$15.5 to $-$16.5 mag, rapid rise times ($\tau_{R} \leq$ 12--15 days),
nebular phase spectra dominated by   prevalent calcium features, 
and are  preferentially located in the outskirts of their host galaxies.   

Adding to this assortment of newly identified transients are  
a  number of peculiar
Type Ia supernovae (SNe~Ia) that probably  
  originated from non-standard thermonuclear explosions. 
Excluding the so-called super-chandrasekhar mass SNe~Ia, 
examples of these  typically faint objects that  show a diverse set of
observational properties include 
SN~2002es \citep{mo12}, SNe~2006bt and 2006ot \citep{foley10b,stritzinger11}, and PTF10ops \citep{maguire11}.
In addition to these  unusual objects are the members of the spectroscopically defined  peculiar 2002cx-like class of SNe~Ia, 
 which have  recently garnered considerable interest within the community. 
SN~2002cx-like or Type~Iax supernovae 
\citep[SNe~Iax, see][and references therein]{foley13} 
exhibit a bizarre set of observational properties  including a considerable range in luminosity ($M_{R}$ $\sim -14$ to $-$19 mag), as well as low values of kinetic energy ($\sim 10^{49}$~erg) and ejected mass (0.15--0.5~$M_{\sun}$). 
Of particular interests is the extreme Type~Iax  
SN~2008ha, which peaks at  
$M_{B}$ $\approx$ $-$14 mag and  has photospheric expansion velocities 
between 4500--5500 km~s$^{-1}$, is the faintest and least energetic 
stripped-envelope SN yet observed \citep{foley09,valenti09,foley10a}.
 
This paper presents  detailed optical and near-infrared (NIR) observations
 of the faint and fast Type Iax~SN~2010ae that as we demonstrate, 
 share many characteristics to SN~2008ha.
 To facilitate a detailed comparison between these two objects,  previously unpublished optical and NIR photometry of SN~2008ha taken by 
 the {\em Carnegie Supernova Project} (CSP; \citealp{hamuy06}) are also presented. 
 To maintain focus on SN~2010ae, details concerning the observations and data reduction of SN~2008ha are deferred to the Appendix.  For an in-depth analysis on SN~2008ha  the reader is referred to papers by \citet{foley09}, \citet{valenti09}, and \citet{foley10a}.

\subsection{Supernova 2010ae}

\begin{figure}[h]
\centering
\includegraphics[width=5.6in]{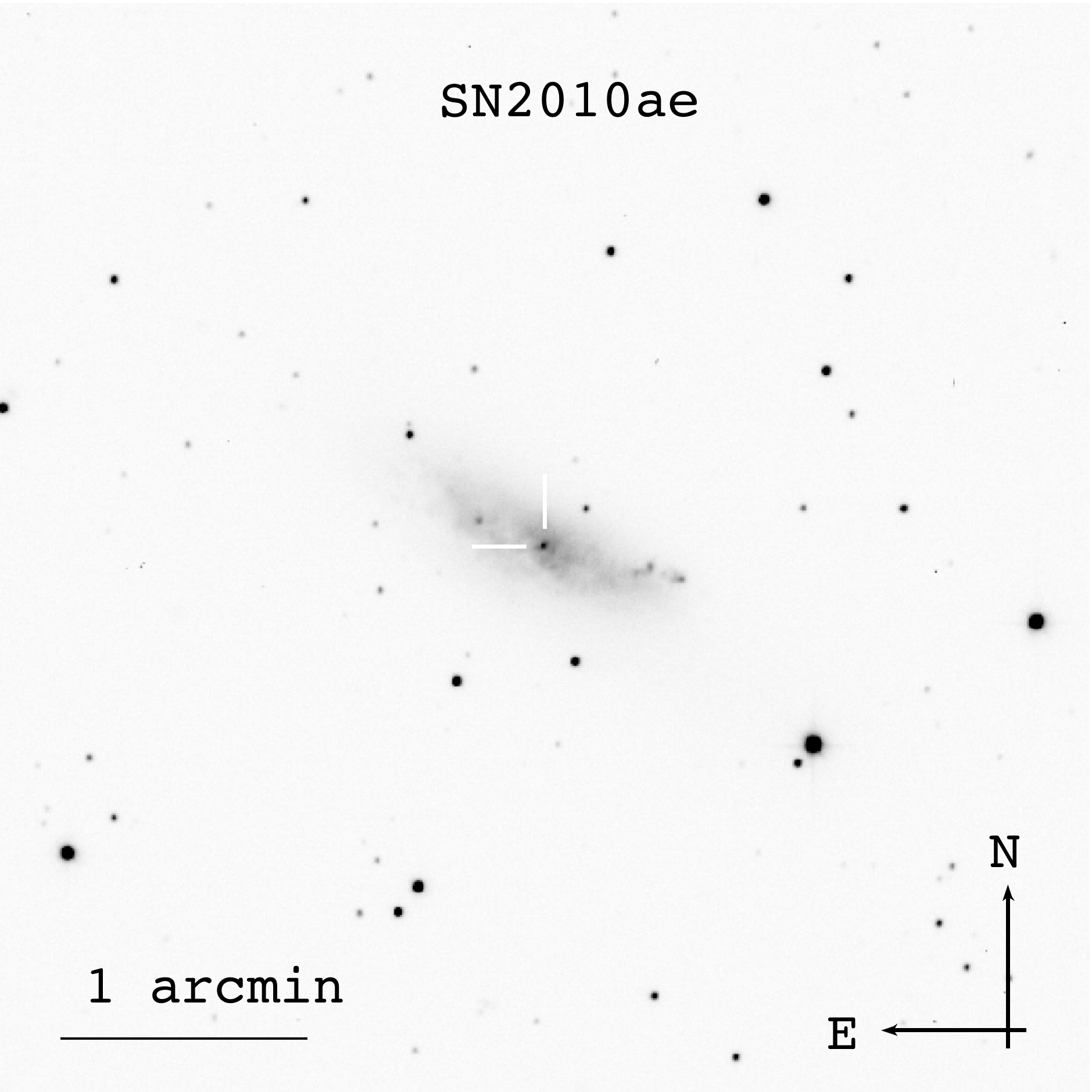}
\caption[]{Swope $V$-band image of the Sb-type peculiar galaxy ESO 162--17, with the position of SN~2010ae indicated.\label{FC}}
\end{figure}

The supernova was discovered in unfiltered  images  by the CHilean Automatic Supernova sEarch (CHASE; \citealp{pignata10}) on 2010 February 22.06 UT.
With J2000.0 coordinates of $\alpha$ $=$ 07$^{\rm h}$15$^{\rm m}$54\fs65 and 
$\delta = -$57$^{\circ}$20$\arcmin$36\farcs9, the location of SN~2010ae
 is less than 1\farcs0 from the center of the type-Sb-peculiar
galaxy ESO 162$-$17 (see Figure~\ref{FC}).  
Previous non-detection  images were obtained
 on 2010 February 10.11 UT and 2010 February 17.11 UT, indicating that the discovery epoch was within a week   after the explosion. 
To determine the detection limit  of these images 1000 artificial stars were distributed randomly within their field, and magnitudes of these stars were  computed with the use of the task {\tt SExtractor}.
From this exercise a robust 3$\sigma$ upper detection limit of 19.2$\pm$0.1 mag was 
computed. However, this upper limit is valid only for the field, whereas at the position of the SN 
the background flux is higher. We then estimated the background noise at the position of the galaxy in the template subtracted pre-discovery search images and compared this to the noise of the field. From this experiment the magnitude limit at the SN position is determined to be  0.33 mag shallower than that of the field. This translates to a final magnitude limit at the position of the SN of 18.9 mag, which is included in the plot of the  broadband light curves  
 presented below (see Figure~\ref{lcs}).

Based on an  initial optical spectrum that exhibited prevalent $\ion{C}{ii}$ $\lambda$6580 absorption, we initially classified SN~2010ae  as a bright SN~Ia \citep{stritzinger10a}.  However, with the addition of  spectra covering an extended wavelength range, it became evident this was a  low-luminosity 2008ha-like SN~Iax around maximum  \citep{stritzinger10b}.
Given its peculiar nature, an intensive  followup campaign was initiated, 
that was made possible through the
collaborative efforts between members of the CSP, 
the {\em Millennium Center for Supernova Science} (MCSS; \citealp{hamuy12}), and the SN ESO Large Program (PI S. Benetti). 

The redshift of ESO 162$-$17  is listed in 
the  NASA/IPAC Extragalactic Database 
(NED) as $z=0.0037$,
which when adopting an $H_{\circ}$ $= 73\pm5$ km s$^{-1}$,
corresponds to a Hubble flow distance (corrected for a Virgo and Great Attractor infall
model) of 12.9$\pm$0.9 Mpc.
 This value is consistent
with the $I$-band Tully-Fisher distance 
of 13.1$\pm$3.5 Mpc ($\mu = 30.58\pm$0.58 mag) as reported by
\citet{springob09}.
In what follows the Tully-Fisher distance is adopted to
set the absolute flux scale of SN~2010ae. 

\section{Host reddening and metallicity estimates}
\label{hostproperties}

The NED galaxy database provides a Milky Way extinction in the direction of  ESO 162$-$17 of
$E(B-V)_{\rm MW} = 0.124\pm0.012$ mag \citep{schlafly11},  that when adopting 
a \citet{fitzpatrick99} reddening law characterized by an $R_{V} = 3.1$,
corresponds to $A_{V} = 0.38$ mag. 

Accurately estimating the reddening of SN~2010ae associated with  
dust external to the Milky Way  is problematic at best. 
Unfortunately, the limited sample of SNe~Iax currently prevents us 
from determining whether or not these objects have standard intrinsic colors \citep{foley13}.
We are therefore largely limited to relying on empirically-derived relations  
 between the equivalent width (EW) of $\ion{Na}{i}~D$ absorption and 
host color excess $E(B-V)_{\rm host}$. These relations  are commonly derived from observations within the Milky Way \citep[e.g.][]{munari97,poznanski12}, 
and provide values with a  $\sim$68\% uncertainty  of the value of $A_V$ \citep{phillips13}.
Nevertheless, as an initial guide to estimating  $E(B-V)_{\rm host}$ we use this relation based on  EW measurements of conspicuous $\ion{Na}{i}~D$ contained within the 
medium-resolution X-shooter spectra (see Section~\ref{earlyspectra}) at the redshift of the host galaxy.
From the series of seven epochs of X-shooter spectra, a Gaussian function was fit to each component, yielding averaged  EW 
values of $\ion{Na}{i}~D1~=~0.74\pm0.06$ \AA\ and 
$\ion{Na}{i}~D2~=~0.58\pm0.06$ \AA.
Combining these averaged values with the empirical relation 
of Poznanski, Prochaska, \& Bloom (2012; see their Eq. 9), which  provides an approximation 
between the EW of  
$\ion{Na}{i}~D$ and the color excess within the Milky Way, we obtain $E(B-V)_{\rm host} =  0.50\pm0.42$ mag.
Adding this to the Galactic component, the total color excess of SN~2010ae is 
 estimated to be  $E(B-V)_{\rm tot} \sim 0.62\pm0.42$ mag (i.e. $A_V = 1.9$ mag). 
This value is consistent with the 
$E(B-V)_{\rm tot} = 0.6$ mag adopted by  \citet{foley13b} for 
a  comparison between  a maximum light spectrum of SN~2010ae  and the normal Type~Ia SN~2011fe. 
We stress that the application of relations between the EW of  
$\ion{Na}{i}~D$ and  color excess are far from quantitative, and 
provide a rather uncertain estimation. 
In what follows, results are therefore provided considering a range  of dust extinction values extending from  the Milky Way component, the combination of the Milky Way and host component derived from the $\ion{Na}{i}~D$ absorption, and  for the intermediate value $E(B-V)_{\rm moderate} = 0.30$ mag. 

Finally, we note that the Balmer decrement was used to provide another 
independent measurement of the total color excess. By extracting a region near the position of the SN from our last epoch optical spectrum, H$\alpha$ to H$\beta$ line fluxes
were measured to give a ratio of 5.7. 
Combining  this ratio with relations between the Balmer decrement and color excess from 
\citet{xiao12} and \citet{levesque10}, we obtain total (host$+$Milky Way) 
color excess values of $E(B-V)_{tot} = 0.59$ mag and $E(B-V)_{tot} = 0.69$ mag, respectively. 
These estimates are consistent with the $E(B-V)_{\rm tot} \sim 0.62$ mag inferred from the $\ion{Na}{i}~D$ absorption.
Below  broadband color curves are examined as a potential avenue for estimating host extinction  (see Section~\ref{sec:lcs}).

To estimate the metallicity of the host near the location of SN~2010ae, we 
turn to line diagnostics derived from conspicuous, narrow host-emission features that are readily measured in the low-resolution optical spectra. 
After careful inspection of the full time series, three epochs of spectra, 
 obtained on 1, 2, and 4 of March 2010 
were deemed to be of high enough quality and  signal-to-noise 
to afford  robust metallicity measurements.

Oxygen abundance metallicities were derived using the 
empirically based N2 and O3N2  calibrations of \citet{pettini04}. 
Gaussian fits  to the H$\alpha$ and $[\ion{N}{ii}]$ $\lambda$$\lambda$6548, 6583 emission lines contained within  the three epochs 
 yield  an averaged local metallicity of $12~+ \log(O/H) = 8.40\pm$0.18~dex, calibrated on the N2 scale.
 Here the quoted uncertainty accounts for both measurement and systematic errors.
For comparison,  the  O3N2 method suggests an averaged local metallicity
of $12~+ \log(O/H) = 8.34\pm$0.14 dex, where again the quoted uncertainty 
includes measurement and systematic errors. 
These  oxygen abundance metallicities 
correspond to 0.52 and 0.44 the known solar 
metallicity of $\sim$8.69 dex \citep{asplund09}, and 
so are consistent with the metallicity of the  LMC.

For comparison,  in the case of SN~2008ha, \citet{foley09} reported  
a  local metallicity of  $12~+ \log(O/H) = 8.16\pm$0.15 dex,  calibrated 
to the N2 and O3N2 scales.
It will be interesting to see in the future if the whole population of SNe~Iax  
occurs  in moderately low metallicity environments, or whether this is limited  only to those objects  located at the extremely faint end of the class.

\section{Observations}
\label{sec:obs}
\subsection{Optical and NIR imaging}

Two months of $g'r'i'z'BV$-band and unfiltered imaging of  SN~2010ae was performed, 
 extending from $-$2 to $+$61 days relative to the epoch of 
 $B$-band maximum (hereafter, $T(B)_{max}$).
The broadband monitoring was obtained as a part of CHASE with the PROMPT telescopes \citep{reichart05}, while
two additional epochs of  imaging was  taken with the Swope 1-meter ($+$SITe3 CCD camera) telescope located at the Las Campanas Observatory (LCO).
Additionally, starting around  maximum and extending for a period of $\sim$ 3 weeks, 
nine epochs of NIR $YJH$-band imaging was also obtained at LCO with the Swope 1-m ($+$RetroCam) and the du Pont ($+$WIRC) 2.5-m telescopes. 
All imaging was processed  in a standard manner following procedures  described in \citet[][and reference therein]{stritzinger11}. 

Photometry of the SN was computed differentially with respect to a  sequence of local stars in the field of the host galaxy. The optical local sequence consists of 22 stars  calibrated with respect to \citet{landolt92} and \citet{smith02} photometric standard fields observed over the course of multiple photometric nights. These standard fields 
therefore provide $g'r'i'z'$ photometry in the AB photometric system and 
$BV$ photometry in the Vega photometry system.
The NIR local sequence consists of 37 stars calibrated using \citet{persson98} standards observed with the Swope, depending on the particular star due to the pointing of the telescope, over the duration of 2 to 8 photometric nights. The $Y$-band local sequence was calibrated relative to the ($J -K_{s}$) relation provided in Hamuy et al. (2006, see Appendix C, Equation (C2)).
Absolute photometry of the local sequence stars is provided in Table~\ref{localsequence}. We note that our entire sequence of unfiltered images was used to compute an 
unfiltered light curve relative to  the local sequence calibrated to  the $R$ band.

Prior to computing  photometry of the SN, galaxy subtraction was performed on all science images in order to remove  significant contamination associated with  
host-galaxy light. Multiple  optical and NIR templates were obtained with the PROMPT and Swope telescopes well after the SN faded. These images were stacked to create deep master templates and implemented to 
subtract the galaxy light at the position of the SN  following the method described by \citet{contreras10}.

\begin{figure}[h]
\centering
\includegraphics[width=5.6in]{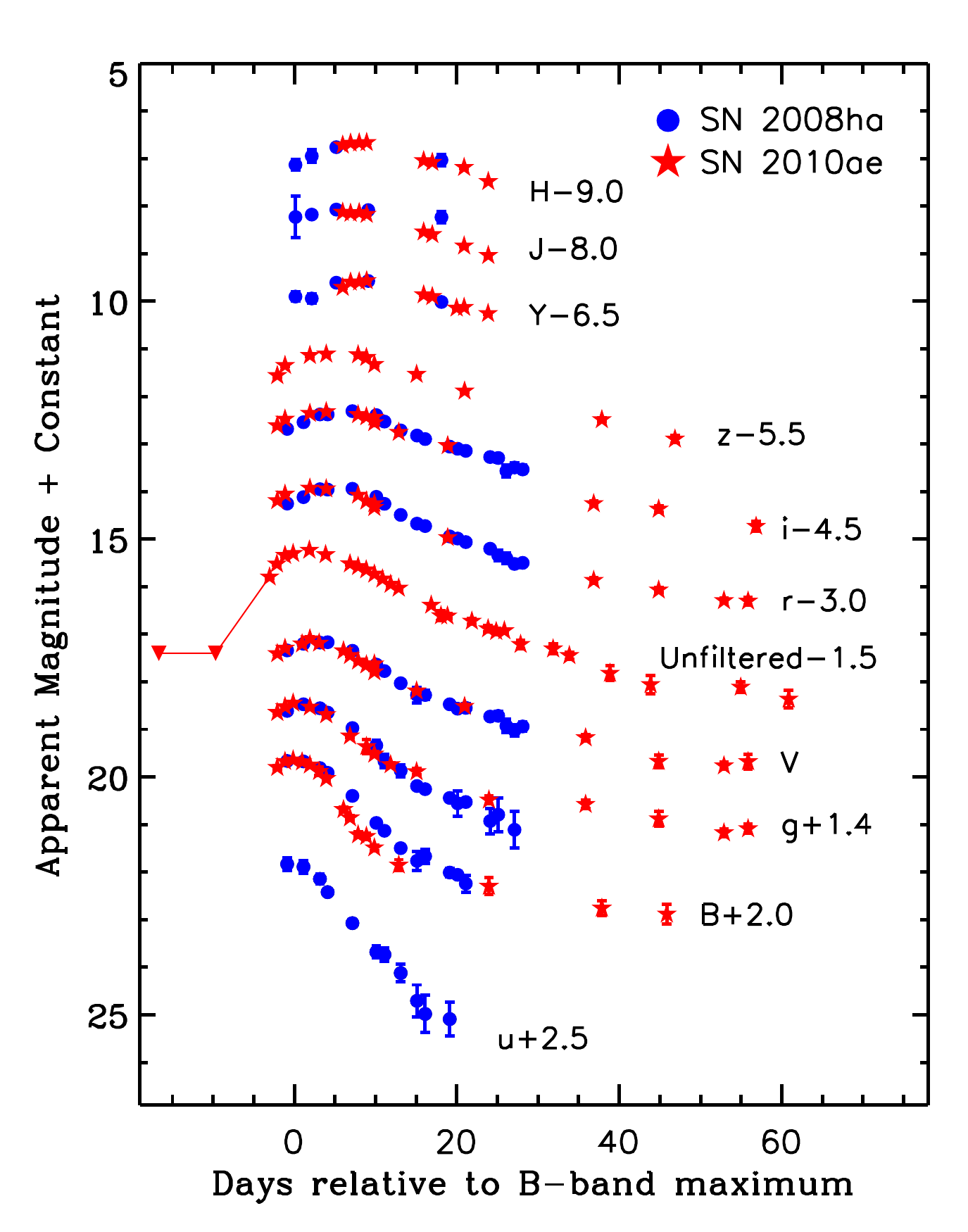}
\caption[]{Optical and NIR light curves of SNe~2008ha (blue dots) and 2010ae (red stars) plotted vs. $T(B_{max})$. The light curves of SN~2008ha have been adjusted in apparent magnitude to match the peak values of SN~2010ae.
Upside-down red triangles are  non-detection upper limits estimated from unfiltered images taken  12 and 5 days prior to discovery.
\label{lcs}}
\end{figure}

The  light curves of SN~2010ae are plotted in 
Figure~\ref{lcs}, and the  corresponding definitive optical and NIR photometry is listed  in Tables~\ref{SN10ae_optphot} and  \ref{SN10ae_nirphot}, respectively.
The optical and NIR light curves of SN~2008ha are also plotted 
in Figure~\ref{lcs} (see Appendix).
For comparative purposes the  light curves of SN~2008ha have been normalized to  match the peak values of SN~2010ae, and are plotted vs. $T(B)_{max}$.

\subsection{Spectroscopy}

With substantial target-of-opportunity (ToO) access on Gemini-South ($+$GMOS) and the VLT ($+$X-Shooter), along with visitor nights at 
the NTT ($+$EFOSC, SOFI), SOAR ($+$GOODMAN), and  du Pont ($+$WFCCD) telescopes, a detailed time series of optical and NIR spectrscopy was obtained for  SN~2010ae.
The resulting early phase time series consists of 21 spectra covering 20 epochs of optical spectroscopy,
extending from $-$2d to $+$57d relative to $T(B)_{max}$,
as well as eight NIR spectra covering seven epochs ranging from  $-$1d to $+$18d.  
Additionally, at late phases a  visual-wavelength spectrum 
was taken with the VLT ($+$FORS2) on  $+$252d.
The journal of spectroscopic observations is presented in 
Table~\ref{specjor}.
 
All low-resolution spectra were reduced in the standard manner using {\tt IRAF}\footnote[13]{The Image REduction and Analysis Facility (IRAF) is distributed by the National Optical Astronomy Observatories, which is operated by the Association of Universities of Research in Astronomy, Inc., under cooperative agreement with the National Science Foundation.} scripts following
the techniques thoroughly described by \citet{hamuy06}. 
To reduce the Gemini-S spectra, we made use of the {\tt IRAF gmos} package following  standard reduction methods.
In the case of the X-Shooter spectra, 
the {\tt esorex} pipeline was utilized to produce rectified 2-D images.
Each 2-D X-Shooter spectrum was then  optimally extracted and flux calibrated 
using the nightly sensitivity function derived from standard star observations. 
Telluric absorption corrections derived from observations of appropriate standards obtained prior to and/or after each set of science exposures were applied to 
each NIR spectrum.
When necessary the fluxing of the time series of 1-D spectra was adjusted  to match 
 the broadband photometric values. In these instances an average value derived from the ratio between the $g'Vr'i'$ synthetic  and  broadband magnitudes was applied to the extracted spectra as a multiplicative  constant. 
 Since fluxing of the Gemini-S spectra was 
 performed through the use of a generic sensitivity function, the adjustments for these particular spectra were at times as large as a factor of 1.5 of the flux. 

\begin{figure}[h]
\centering
\includegraphics[width=5.6in]{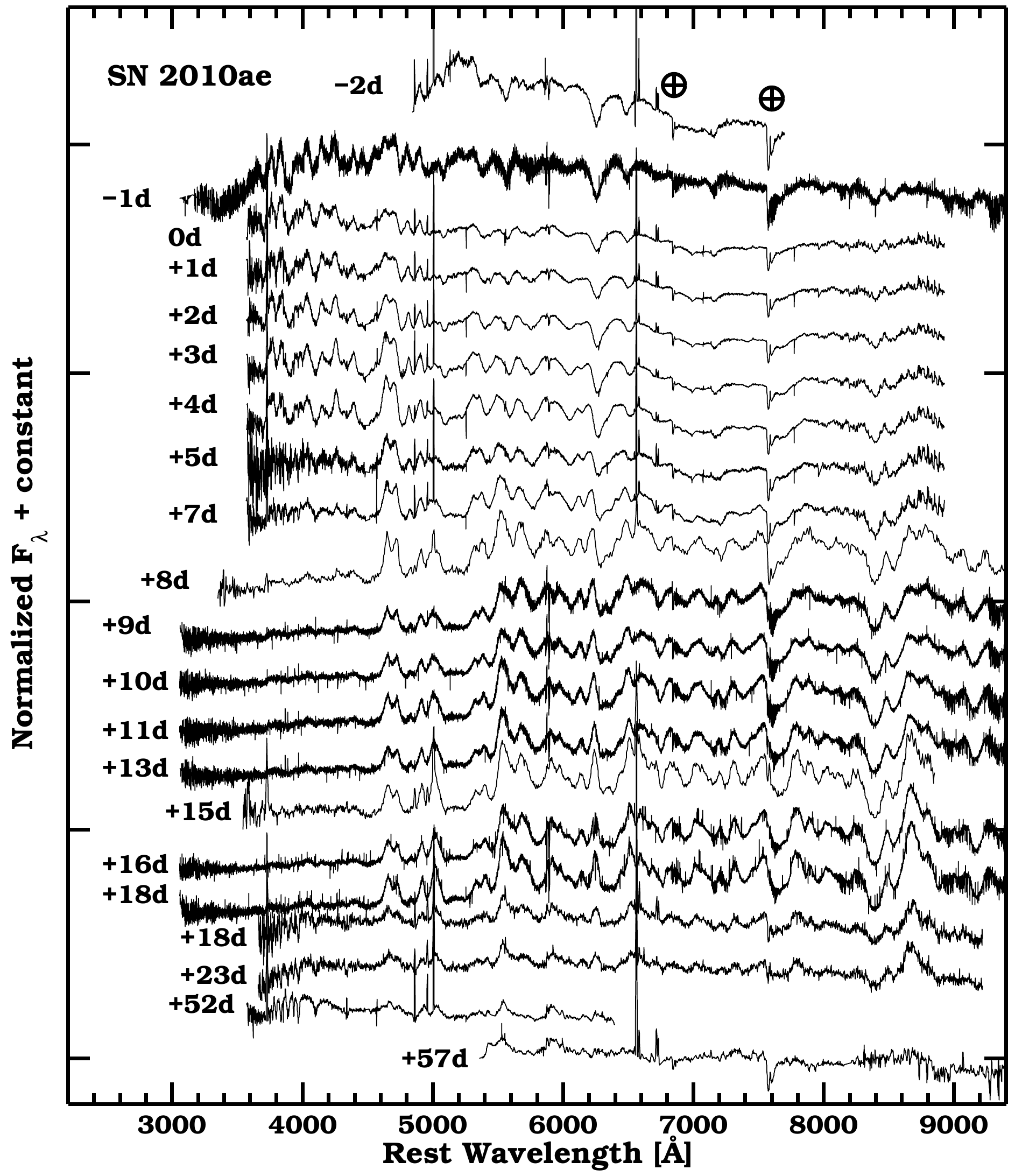}
\caption[]{Optical spectroscopic time series of SN~2010ae. 
Each spectrum has been normalized and corrected to the rest frame of the host galaxy adopting the redshift $z = 0.0037$.
The labels on the left indicate the rounded epoch relative to $T(B)_{max}$.
We note for clarity each X-Shooter spectrum has been moderately smoothed.  Prominent telluric features are indicated with an Earth symbol. 
\label{optspec}}
\end{figure}

\begin{figure}[h]
\centering
\includegraphics[width=5.6in]{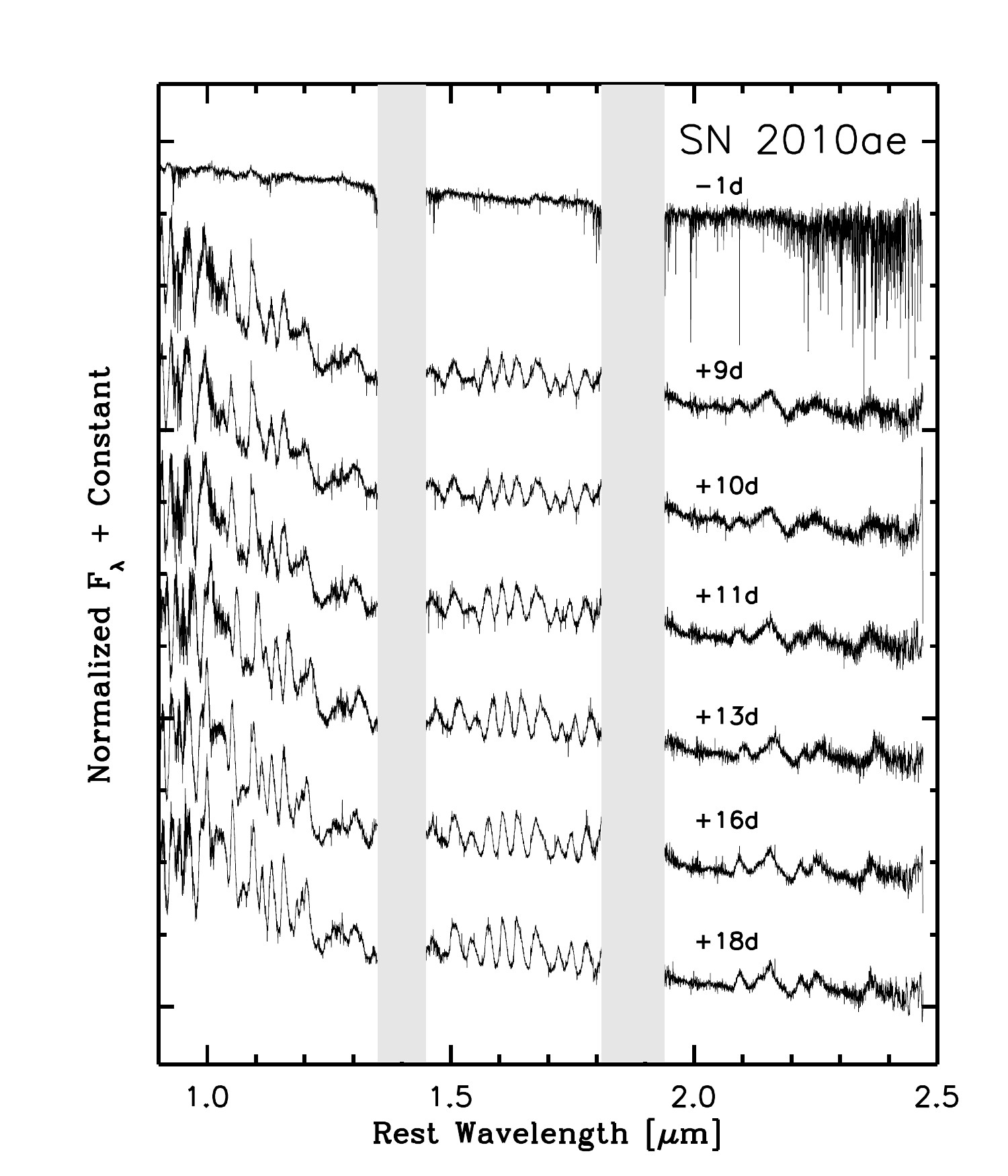}
\caption[]{NIR spectroscopy of SN~2010ae taken with the VLT equipped with X-Shooter. 
Each spectrum has been smoothed, normalized, and corrected to the rest wavelength of the host adopting the redshift $z = 0.0037$. 
The labels on the right indicate the rounded epoch relative to $T(B)_{max}$.
Vertical  gray bands mask  the most prevalent telluric regions.
\label{nirspec}}
\end{figure}

The final optical and NIR spectroscopic sequences of SN~2010ae are plotted 
in Figures~\ref{optspec} and~\ref{nirspec}, respectively, and 
the late-phase optical spectrum of SN~2010ae is shown in 
Figure~\ref{latespectra},
compared to a similar epoch  spectrum of the Type~Iax SN~2002cx \citep{jha06}.
 
\section{Light Curve Analysis}
\label{sec:analysis}

\subsection{Optical and near-IR Light Curves}
\label{sec:lcs}

The photometry of SNe~2008ha and 2010ae plotted in Figure~\ref{lcs} reveal 
similar bell-shapped light curves, characterized by a fast rise to maximum followed by a subsequent decay, and yield no evidence for a secondary maximum.
Basic light curve parameters of these two objects were measured from 
 Guassian process  functional fits to the photometry.
Table~\ref{lcpar} lists the key observables including the time of maximum,  apparent and absolute peak magnitudes, and values of the decline-rate parameter, $\Delta$m$_{15}$.
Here $\Delta$m$_{15}$ is 
 defined as the magnitude change  from the time of maximum brightness
to 15 days later. In the case of normal SNe~Ia, $\Delta$m$_{15}$ is known
to correlate with the peak absolute luminosity in such a way that more luminous objects exhibit smaller  $\Delta$m$_{15}$  values \citep{phillips93}.
However, in the case of SNe~Iax this relationship is known to exhibit significant scatter \citep{foley13}.   
The quoted uncertainty of the light curve fit parameters summarized in 
Table~\ref{lcpar} are robust estimations derived from Monte Carlo simulations, while the associated uncertainty of the absolute magnitudes  account for both errors in the fit and the adopted distance.
Given our inability to accurately estimate the host extinction, a range in absolute magnitude is given for each passband assuming  the Galactic component and 
$E(B-V)_{tot} = 0.62$ mag.  We note that in order to  compute the
peak absolute magnitudes of SN~2008ha 
a Galactic reddening value of $E(B-V)_{MW} = 0.07$ mag was used and no host reddening was adopted.

Functional fits to the  light curve fits of both SN indicate the following general trends: 
(i) the bluer the bandpass, the earlier in time maximum light is reached, with a $\approx9$-day delay between the time of $B$-band and $H$-band maxima and,
(ii) the bluer the filter, the faster the light curve evolves as parameterized by 
the decline-rate $\Delta$m$_{15}$.

At peak brightness SNe~2008ha and 2010ae reached 
$M_{B} = -13.79\pm0.14$ mag and,  depending on the adopted 
extinction value, $-13.44\pm0.54 \gtrsim M_{B} \gtrsim -15.47\pm0.54$ mag, respectively.
Even when adopting  an $A_{B} = 2.5$ mag, 
SN~2010ae comfortably  sits as the  second least luminous SN~Iax yet observed, and
if only Galactic extinction is adopted,  is  0.3 mag less luminous than SN~2008ha.
The $H$-band light curves, which are less susceptible to dust extinction imply 
SN~2008ha is at most $\sim$0.8 mag brighter than SN~2010ae (see Table~\ref{lcpar}).
Interestingly,
 SN~2010ae  exhibits a faster $B$-band light curve decline with  $\Delta$$m_{15}(B) = 2.43\pm0.11$ mag
as compared to SN~2008ha with $\Delta$$m_{15}(B) = 2.03\pm0.20$ mag, however, SN~2008ha has marginally higher decline rates in the redder bands (see Table~\ref{lcpar}).

\begin{figure}[h]
\centering
\includegraphics[width=6.0in]{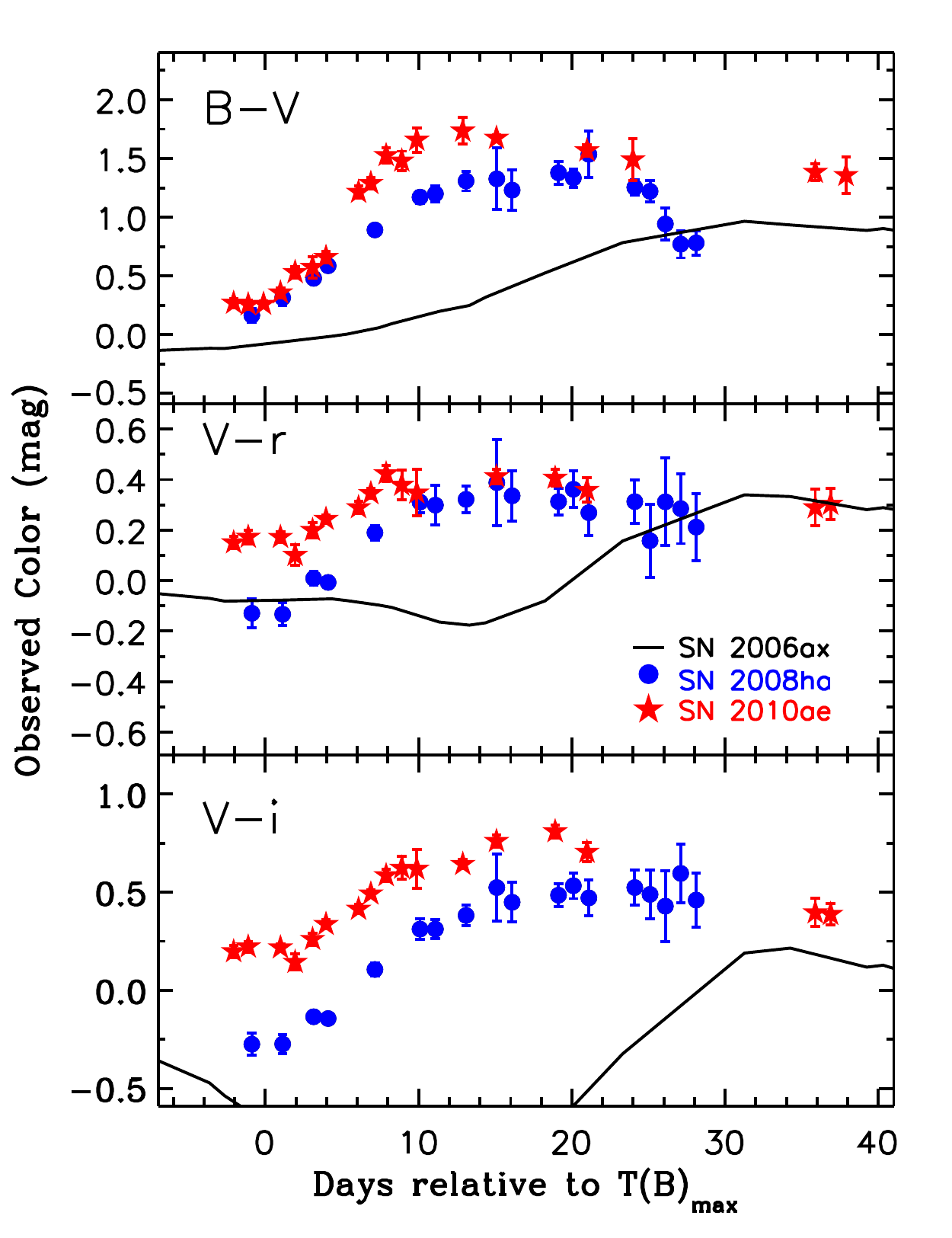}
\caption[]{($B-V$), ($V-r$), and ($V-i$) color curves of the low-luminosity Type~Iax SNe~2008ha (blue dots) and 2010ae (red stars), compared to the normal and unreddened
 Type Ia SN~2006ax (solid line). The color curves have been corrected for Milky Way extinction.\label{color}}
 \end{figure}

We now compare  optical colors  of SNe~2008ha and 2010ae.
Color curves normally tend to track the photospheric temperature evolution, and in the case of {\em some} SN types, can provide constraints on dust extinction. 
The ($B-V$), ($V-r$) and ($V-i$) color curves of 
SNe~2008ha and 2010ae are plotted in Figure~\ref{color}, corrected only for Galactic reddening. 
For comparison,  the ($B-V$) and  ($V-r$) color curves of the normal Type~Ia 
SN~2006ax \citep{contreras10}, are also plotted in Figure~\ref{color}, and reveal a noticeably  different morphology.
At the earliest epochs the colors of SNe~2008ha and 2010ae 
are at their bluest value, 
 corresponding to the phase when  the photospheric temperature 
is at its highest value. 
 As the ejecta expand and cool the colors evolve towards the red, 
 reaching a maximum value between 15$-$20 past $T(B)_{max}$, 
 whereupon the colors appear to level off, or in the case of the ($B-V$) color curve of SN~2008ha evolves back towards the blue.

 Comparing the color curves between the two objects, it is evident that at maximum brightness they exhibit nearly identical ($B-V$) colors, however, by $+$10d SN~2010ae appears $\approx$ 0.6 mag redder than SN~2008ha. 
 Interestingly, this is consistent with the estimated upper limit on 
 $E(B-V)_{\rm tot} = 0.62$ mag.
 However, inspection of the ($V-r$) and ($V-i$) colors at the same epoch
 reveals that SN~2010ae is redder than SN~2008ha by about half of what is expected for $E(B-V)_{\rm tot} = 0.62$ mag. This highlights the complexities of attempting to separate intrinsic colors  from effects associated with dust reddening of SNe~Iax, and further progress will certainly require an expanded sample to determine if it is at all possible to disentangle these parameters. 
 
\subsection{SEDs, UVOIR light curves and light curve modeling}
\label{sectionuvoir}

The comprehensive optical and NIR photometry allow for the 
construction of  broadband spectral energy distributions (SEDs) and 
UVOIR (Ultra-Violet-Optical-InfaRed) bolometric light curves.
To begin, the NIR light curves of SNe~2008ha and 2010ae were 
interpolated so NIR magnitude could be measured on dates corresponding
to  the optical-band observations. 
Additionally, the NIR light curves also required extrapolation as their temporal coverage is not as complete as in the optical.
This was accomplished by computing the   ($i-Y$), ($Y-J$), and ($J-H$) colors  at the first and last observed NIR epochs. These colors were then used to extrapolate  the NIR light curves to all phases covered by the optical broadband observations. 
In the case of SN~2008ha its  $u$-band light curve was also extended in time adopting the ($u-B$) color computed from photometry obtained on  the last epoch of  $u$-band observations. 
Additionally, to extend the time coverage of SN~2008ha's UVOIR light curves beyond $+$28d, we utilized  published broadband photometry presented by \citet{foley09} and \cite{valenti09}. 
All of the photometry was  next corrected 
for extinction and converted to flux at the effective wavelength of each passband, allowing for the construction of the SEDs. 

\begin{figure}[h]
\centering
\includegraphics[width=5.6in]{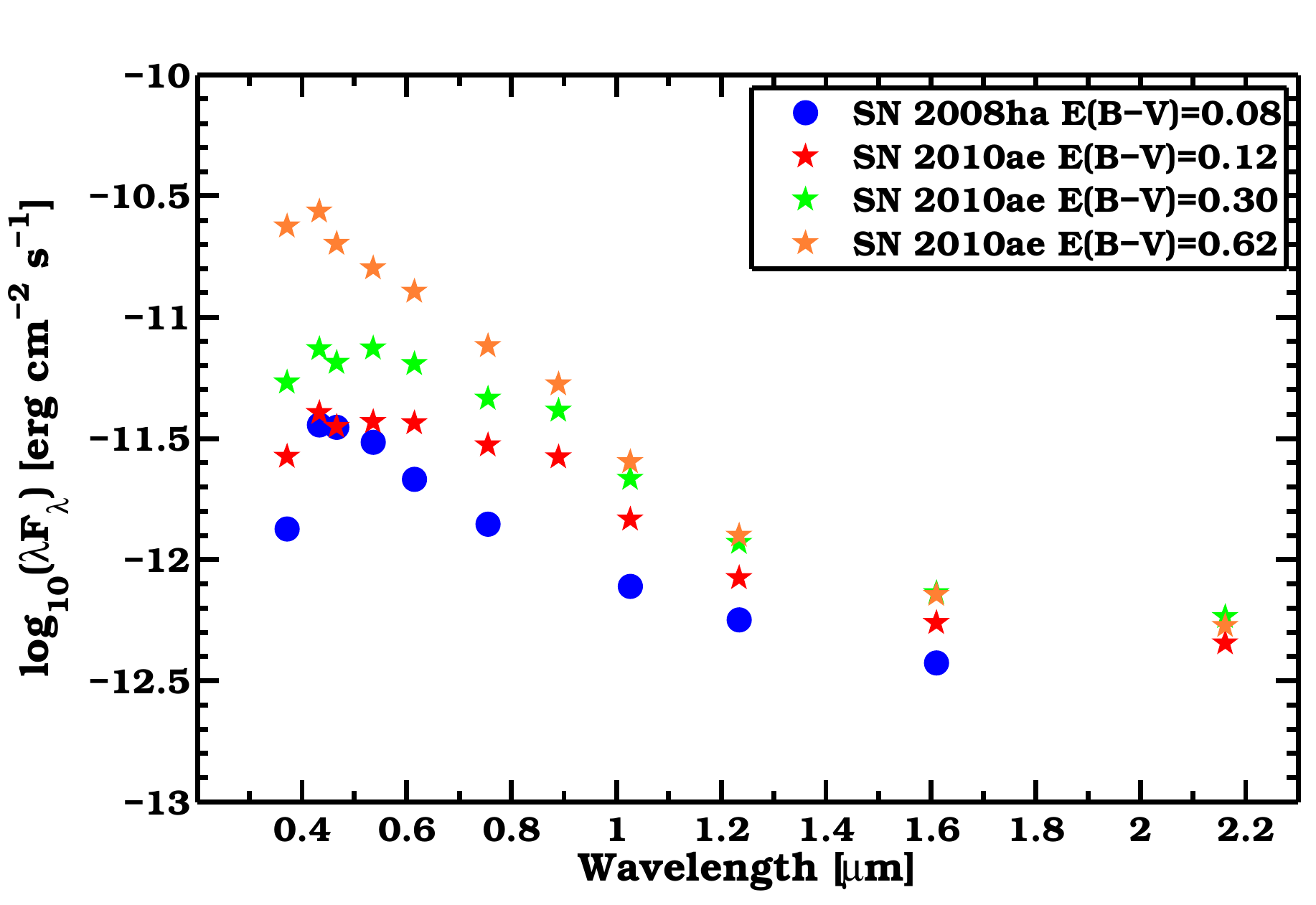}
\caption[]{SEDs of SN 2008ha (dots) and SN~2010ae  (stars) 
at maximum light. SEDs of SN~2010ae are plotted for three reddening values, 
corresponding  to the  Galactic component  (red stars) 
the combined 
Galactic plus host component as estimated from the equivalent width of the host's $\ion{Na}{i}~D$ absorption features, and an intermediate value of these two estimates
(see Section~\ref{hostproperties}).
The SED of SN~2008ha is constructed from the $u'Bg'Vr'i'YJH$ broadband
observations corrected for Galactic reddening. The SEDs  of SN~2010ae
are constructed from measurements obtained with the $Bg'Vr'i'z'YJH$ passbands, while 
the $u'$- and $K_s$-band flux points correspond to measurements provided by 
the first X-Shooter spectrum. 
\label{sed}}
\end{figure}

Plotted in Figure~\ref{sed} are the  SEDs of SNe~2008ha and 2010ae at maximum light. In the case of SN~2010ae, SEDs are shown for three color excess values. 
Clearly at this phase (and later) most of the  flux is emitted at optical wavelengths,
while SN~2010ae shows more flux at red wavelengths than does SN~2008ha.
The effect of adopting the high extinction value on the SED of SN~2010ae leads to a larger departure than both the Galactic only extinction value  and 
SN~2008ha. If one were to assume that these two SN have similar intrinsic colors,
this would then suggest that the $E(B-V)_{\rm host}$ value  based on the EW of $\ion{Na}{i}~D$ absorption provides an over estimation.

\begin{figure}[h]
\centering
\includegraphics[width=5.6in]{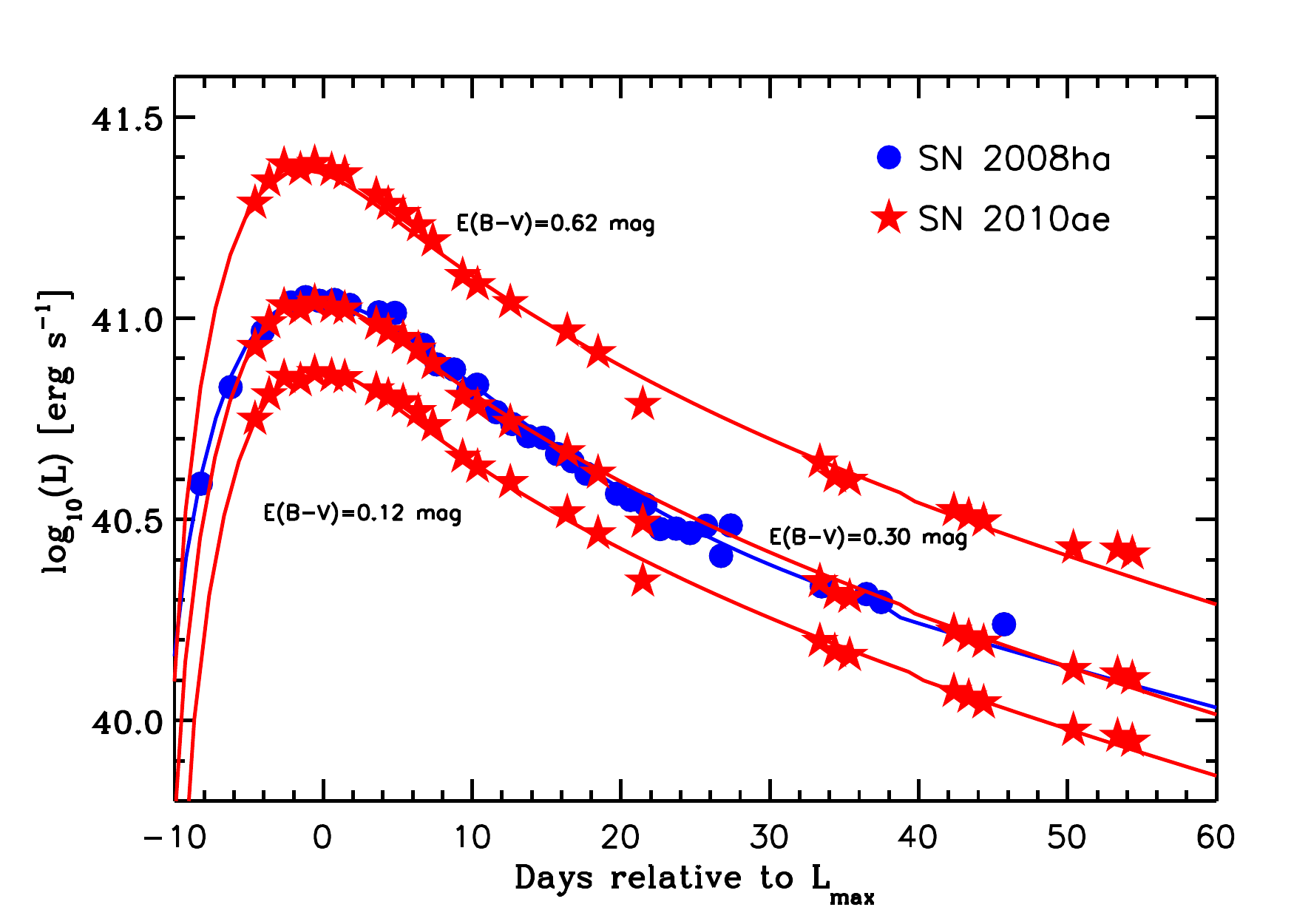}
\caption[]{Comparison of the UVOIR light curves of SNe~2008ha and 2010ae.
To ensure consistency each UVOIR light curve was constructed through the direct integration of flux between 
the $B$ and $H$ bands.
Given the uncertainty in the reddening estimate of SN~2010ae,  UVOIR light curves are plotted assuming $E(B-V)_{\rm MW} = 0.12$ mag, $E(B-V)_{\rm intermediate}$ $=$ 0.30 mag and $E(B-V)_{\rm tot} = 0.62$ mag
(see Section~\ref{hostproperties}). Over-plotted  the UVOIR light curves as solid lines are  model fits,  from which the values of $M_{Ni}$, $M_{ej}$, and $E_{K}$ are estimated (see Section~\ref{sectionuvoir}).\label{uvoir}}
\end{figure}

To compute the UVOIR light curves the full time series of SEDs were integrated over flux  from $B$ to $H$ bands using a simple trapezoidal  technique, 
and assuming zero flux beyond the extremes of the wavelength range. 
The results  are shown in Figure~\ref{uvoir}, where for SN~2010ae  UVOIR light curves  are presented for values of  $E(B-V)_{MW} = 0.12$ mag,  $E(B-V)_{moderate} = 0.30$ mag, and $E(B-V)_{tot} = 0.62$ mag.
Like  the individual absolute magnitude light curves, SN~2010ae is found to be, depending on the adopted color excess value, marginally fainter, consistent with or brighter than SN~2008ha.

Key explosion parameters of SNe~2008ha and 2010ae were estimated 
through model fits to the UVOIR light curves.
The model is based on  analytical solutions to 
Arnett's equations \citep{arnett82}, and provides estimates 
of ejecta mass ($M_{ej}$), the radioactive $^{56}$Ni content ($M_{Ni}$),  and the  kinetic energy ($E_{k}$) of the explosion. 
 The  model fits follow the methodology presented by 
 Valenti et al. (2008; see their Appendix A), and 
 relies upon a number of  underlying assumptions 
  including: homologous expansion of the ejecta, spherical symmetry, no appreciable mixing of $^{56}$Ni, constant optical opacity (in this case $\kappa$$_{opt} =$ 0.1 cm$^{2}$~g$^{-1}$), and the diffusion approximation for photons.
 An important input parameter used to compute a model light curve is an expansion 
 velocity ($v_{ph}$) of the ejecta. 
 We opted to compute a grid of models that encompasses a range of velocities 
 bracketing the value of the photospheric velocity  inferred from {\tt SYNAPPS} \citep{thomas11} synthetic spectral fits to our near maximum light optical spectra of SNe~2008ha and 2010ae (see below).
 Our {\tt SYNAPPS} analysis of the near maximum light spectra of SNe~2008ha and 2010ae provides photospheric expansion velocities that range between  4500 $< v_{ph} <$ 5500 km~s$^{-1}$ and 5000 $< v_{ph} < 6000$ km~s$^{-1}$, respectively.
 
  Final best-fit modeled light curves of SNe~2008ha and 2010ae  are  plotted as solid lines in Figure~\ref{uvoir}. 
 The corresponding explosion parameters computed to provide  a good
 match to the UVOIR light curve of  SN~2008ha are $M_{Ni}$ $=$ 0.004 $M_{\sun}$, $M_{ej}$ $\sim$ 0.4--0.5 $M_{\sun}$ and  $E_{k}$ $\sim$ 0.08--0.18$\times$10$^{51}$ erg.
  In the case  of SN~2010ae,  because of the large uncertainty associated with the
   host extinction, modeled parameters were derived for each of the UVOIR light curves shown in Figure~\ref{uvoir}.  
   This analysis provides low, medium, and high adopted extinction 
   modeled parameters of  (i) $M_{Ni}$ $=$ 0.003, 0.004 and 0.007 $M_{\sun}$;
   (ii)  $M_{ej}$ $\sim$ 0.45--0.60, 0.35--0.50, and 0.30--0.50 $M_{\sun}$; and (iii) 
 $E_{k}$ $\sim$ 0.09--0.30$\times$10$^{51}$,  0.06--0.20$\times$10$^{51}$,
 and  0.04--0.26$\times$10$^{51}$ erg. 
 In summary, when comparing these two faint and fast transients, depending on the adopted host extinction value, we find that SN~2010ae 
 produced a similar amount of  $M_{Ni}$, and marginally higher 
 values of $M_{ej}$ and $E_{k}$ than SN~2008ha.

 \section{Spectroscopic analysis}
\subsection{Early phase spectroscopy}
\label{earlyspectra}

The spectroscopic time series of SN~2010ae   (see Figures~\ref{optspec} and \ref{nirspec}) represents among the most complete temporal and wavelength coverage yet obtained for an SN~Iax, allowing for the opportunity to gain 
 new insight into the nature of this class of transients. 
The visual-wavelength time series exhibits rich structure characterized by numerous
low-velocity P-Cygni spectral features formed by both  
 intermediate-mass elements (IMEs)  and Fe-group elements. 
As the SN evolves through maximum, the strength of many of the IMEs 
decreases, even becoming in some cases un-identifiable (e.g. $\ion{C}{i}$), while simultaneously, features  associated with Fe-group elements begin to  dominate. 
At NIR wavelengths the $-$1d spectrum exhibits a  rather smooth, 
featureless continuum superposed by a handful of notches.  
In the next observed  spectrum ($+$9d)  a dramatic transformation occurs, 
revealing a significant number of  prominent low-velocity emission features.  
 The spectral evolution   is reminiscent of that observed in normal SNe~Ia \citep[e.g.][]{hsiao13}; however, given  the extremely low kinetic energy of SN~2010ae, 
its  NIR spectrum exhibits numerous features that emerge relatively rapidly after maximum.    
    
 Close examination of the high signal-to-noise spectra of SN~2010ae indicates the presence of a multitude of host-galaxy emission lines associated with 
a diffuse  $\ion{H}{ii}$ region. Discernible emission features include: 
$[\ion{O}{ii}]~\lambda$3727, $[\ion{Ne}{iii}]$~$\lambda$3969,  
$[\ion{O}{iii}]~\lambda\lambda$4959, 5007, $[\ion{N}{ii}]~\lambda\lambda$6548, 6583,
 $[\ion{S}{ii}]~\lambda$6716, and $[\ion{Ar}{iii}]~\lambda$7136, as well as 
 Balmer lines at $\lambda\lambda$3835, 3889, 3970, 4102, 4340, 4861, 6563, and
$\ion{He}{i}$ $\lambda\lambda$5876, 6678, 7065 emission features.
From a detailed inspection of the 2-D spectra, we can conclude that the extended 
$\ion{H}{ii}$ region emission is in the vicinity  of the SN; however, the local peak observed in 
$H\alpha$ is found not to be coincident with the position of the SN.

In addition to the nebular emission lines, an excess of flux blue-wards of 
$\sim$ 4500 \AA\ is evident in the $+$52d and $+$248d (see below) spectra, 
most likely due to incomplete background subtraction. Under closer scrutiny, 
the absorption features contained within the blue portion of the $+$52d spectrum 
closely resemble those of an E$+$A post-starburst galaxy. 
From this we infer that SN~2010ae likely exploded in an environment which 
underwent an episode of star formation within the last $\sim$1 Gyr

 Returning to the SN features, in order to 
 understand which ions are responsible  for the 
 numerous absorption and emission features that characterize 
 the spectroscopic time series of SN~2010ae, we turn to the  parameterized 
 spectral synthesis code {\tt SYNAPPS} \citep{thomas11}.
 Based on the workings of {\tt SYNOW}  \citep{fisher00,branch09}, 
{\tt SYNAPPS} is an enhanced and automated spectral synthesis code 
that relies on a number of underlying assumptions \citep[for complete details see][and references therein]{thomas11}, and therefore its results should be approached 
with caution.
However, it does provide a fast and effective guide for line 
identification for a variety of  SN types, and is quite useful  for studying  
transient objects that are poorly understood. 
The input parameters required to compute a synthetic spectrum consist of a
black-body temperature ($T_{BB}$), an e-folding  velocity ($v_{e}$) 
for the exponentially declining optical depth distribution, and a list
of ions. Each ion requires an additional set of input parameters, including 
 an excitation temperature ($T_{exc}$), an optical depth ($\tau$), and
a photospheric velocity ($v_{ph}$) of the opacity distribution.

\begin{figure}[h]
\centering
\includegraphics[width=5.6in]{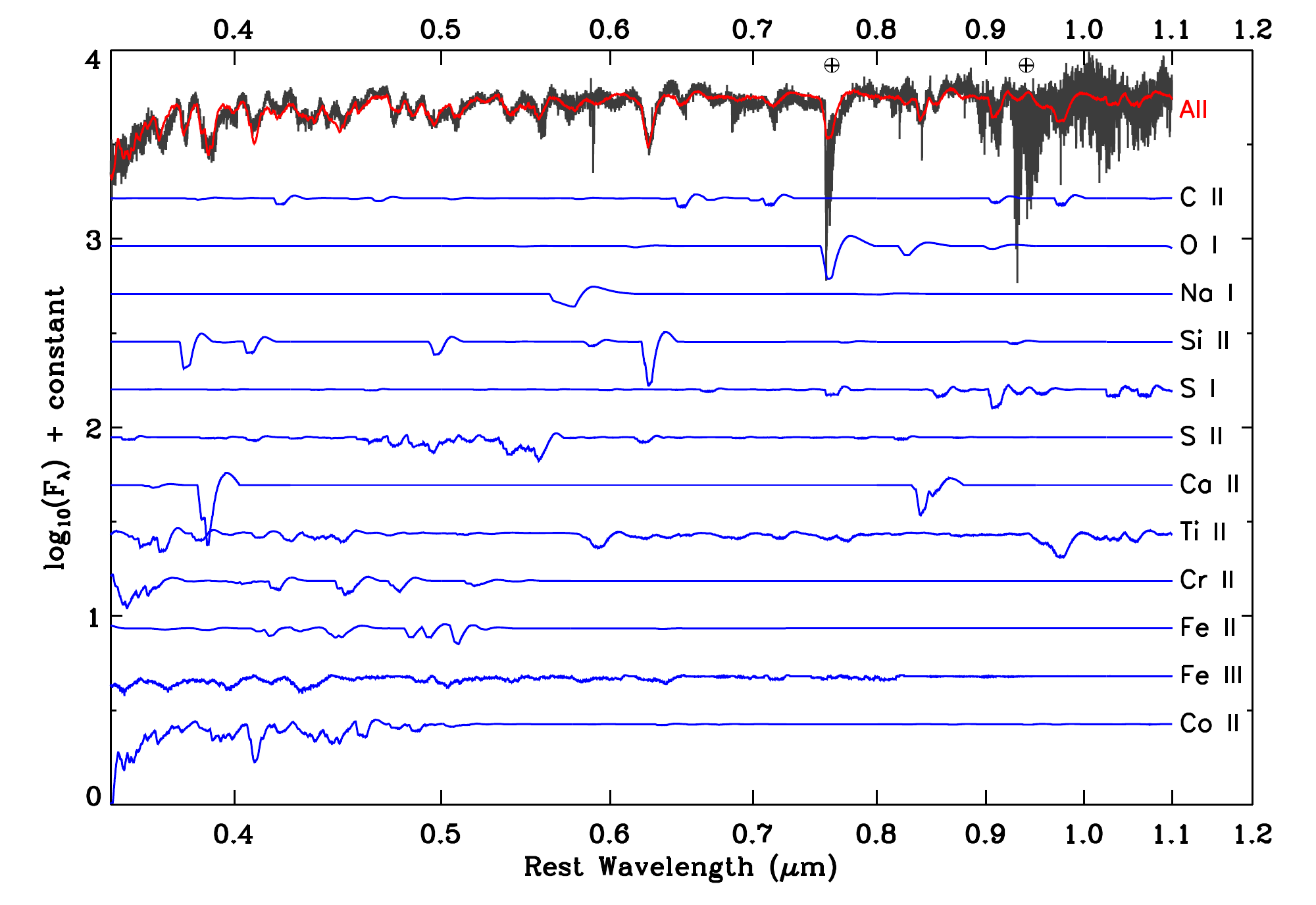}
\caption[]{Visual-wavelength $-$1d X-Shooter spectrum of SN~2010ae (black),
 compared to 
the best-fit  {\tt SYNAPPS} synthetic  spectrum (red). 
The various ions included in the model calculations are also plotted.
The observed spectrum has been flattened and sigma-clipped. 
Prevalent telluric regions are indicated with an Earth symbol. 
\label{synapps1}}
\end{figure}

To identify the most pertinent spectral features  {\tt SYNAPPS} models were computed  for the  $-$1d and $+$18d spectra.
Given the lack of absorption features in the $-$1d NIR spectrum,  a fit at this 
epoch was limited to only optical wavelengths. 
 When computing the   {\tt SYNAPPS} spectrum an extended set of ions 
was used in the calculation including: $\ion{C}{i}$, $\ion{C}{ii}$, $\ion{O}{i}$, $\ion{Na}{i}$, 
$\ion{Mg}{i}$, $\ion{Mg}{ii}$,
$\ion{Si}{ii}$, $\ion{Si}{iii}$, $\ion{S}{i}$,  $\ion{S}{ii}$, $\ion{Ca}{ii}$, $\ion{Sc}{II}$, $\ion{Ti}{ii}$, $\ion{Cr}{ii}$, $\ion{Fe}{ii}$, $\ion{Fe}{iii}$, and $\ion{Co}{ii}$.
The resulting best-fit synthetic spectrum is compared to the $-$1d spectrum in 
Figure~\ref{synapps1}.
Here the synthetic spectrum was computed with $v_{ph} = 5960$ km~s$^{-1}$, and as indicated in the figure a  subset of 12 of the above IMEs and Fe-group ions, which 
provide plausible contributions to the observed spectral features. 
Ions of IMEs  account for numerous spectral features throughout the optical spectrum, while  Fe-group ions are largely responsible for forming a multitude of features that dot the blue end of the spectrum
\citep[for a slightly alternative SYNOW spectrum, see][]{foley13b}.
 
\begin{figure}[h]
\begin{center}$
\begin{array}{cc}
\includegraphics[scale=0.75]{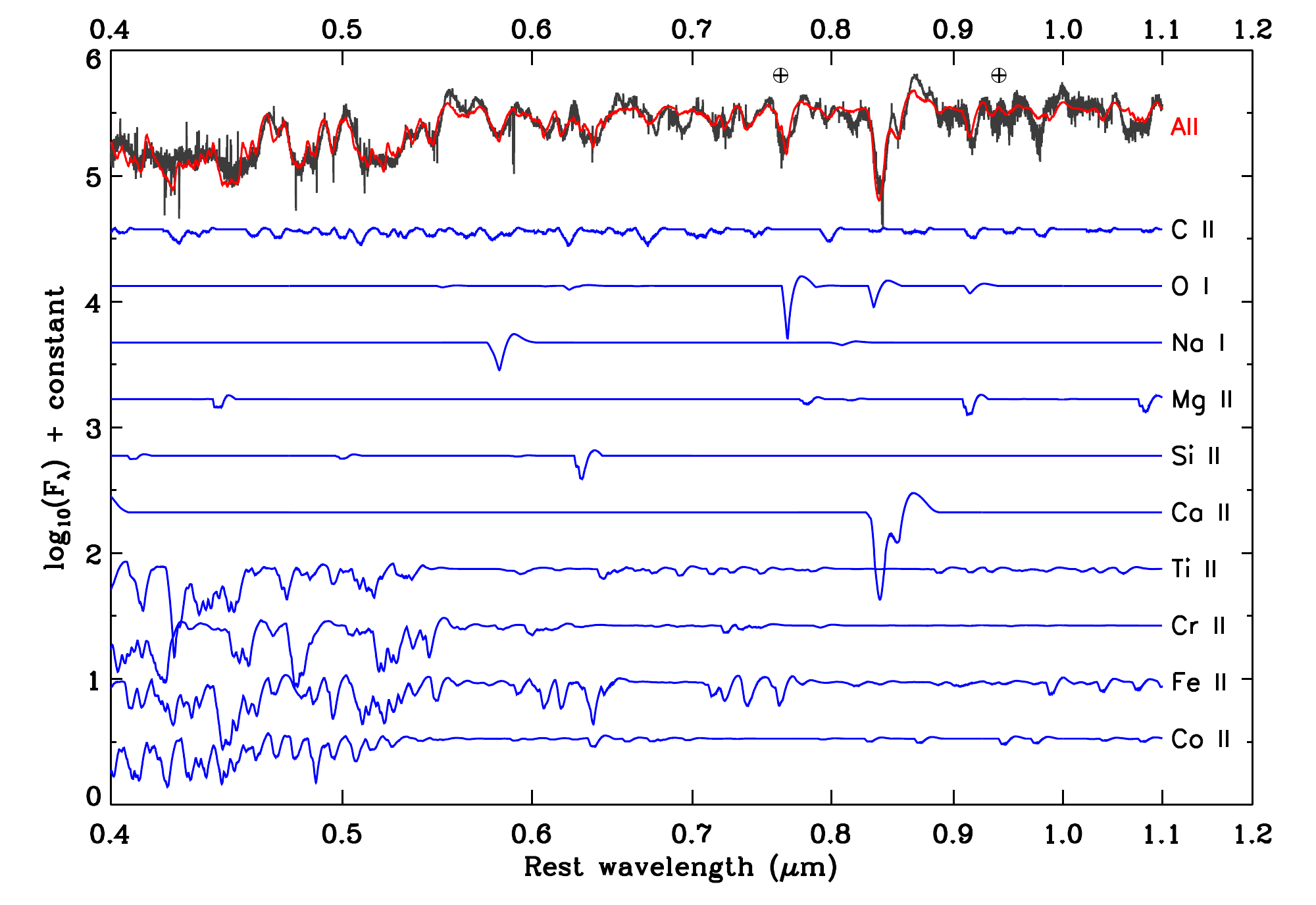} \\
\includegraphics[scale=0.75]{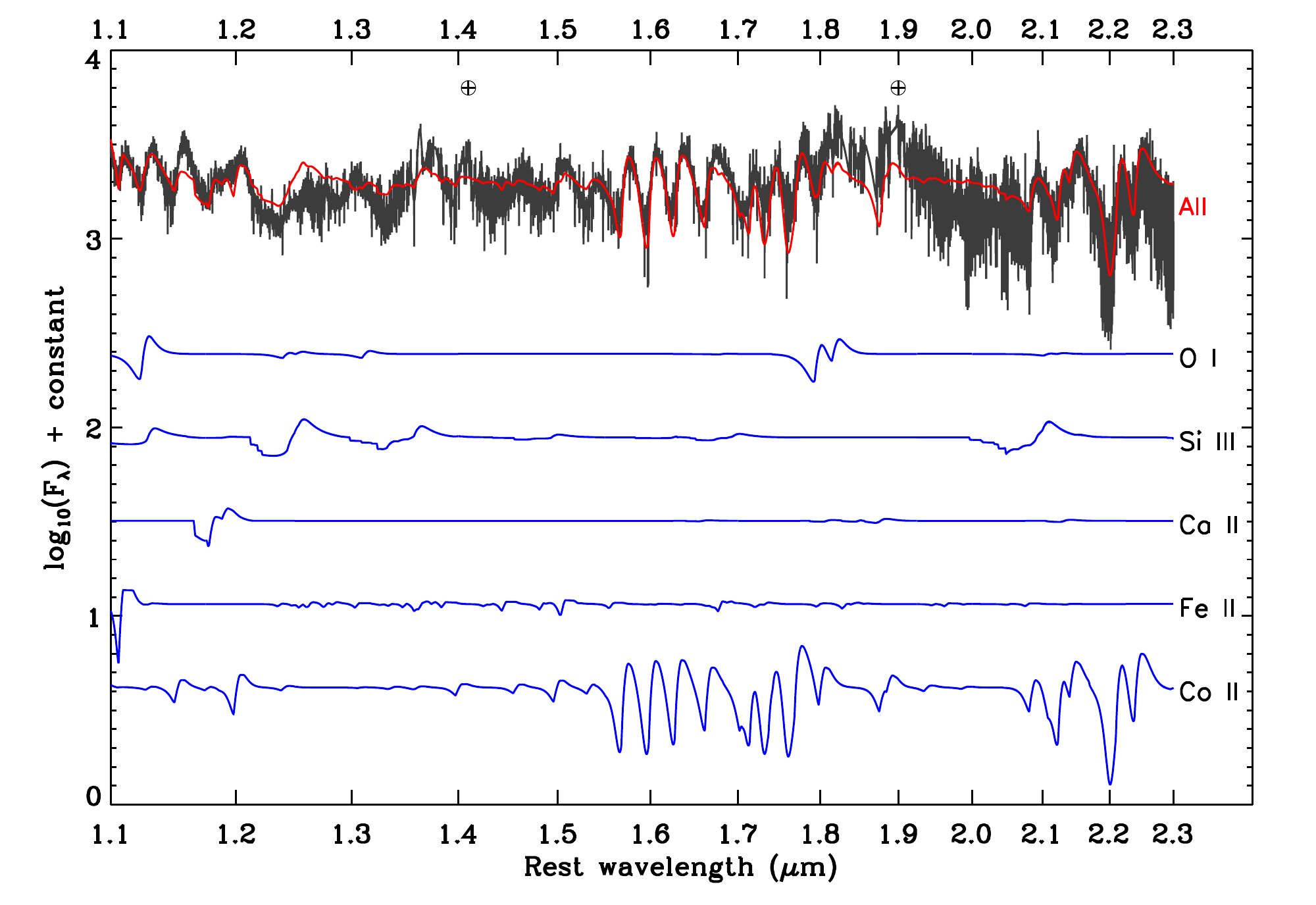} \\
\end{array}$
\end{center}
\caption{Visual-wavelength ({\em top}) and NIR ({\em bottom}): 
$+$18d X-Shooter spectrum of SN~2010ae (black), compared to 
the best-fit  {\tt SYNAPPS} synthetic  spectrum (red). 
The various ions included in the model calculations are also plotted.
As discussed in Section~\ref{earlyspectra} the synthetic spectra shown here 
were computed using  a different input set of 
ions and values of $v_{phot}$. 
The observed spectrum has been flattened and sigma-clipped. 
Prevalent telluric regions in the optical are indicated with an Earth symbol.
\label{synapps2}}
\end{figure}

To obtain a synthetic spectrum that provides a reasonable match to the extended $+$18 day X-Shooter spectrum, it 
was found that a  subset of ions characterized by different $v_{ph}$ values 
provides the best fits to the observed spectrum.
Figure~\ref{synapps2} displays the resulting  best-fit {\tt SYNAPPS} synthetic spectrum, where the optical and NIR spectra 
are shown in the top and bottom panels, respectively.
At optical wavelengths a decent synthetic spectrum fit is obtained 
with $v_{ph} = 4200$  km~s$^{-1}$, and the same ions as indicated in 
Figure~\ref{synapps1}, except excluding $\ion{S}{i}, \ion{S}{ii}$ and $\ion{Fe}{iii}$, and 
including $\ion{Mg}{ii}$.  
Instead, the NIR spectrum is found to be  described 
with $v_{ph} = 1700$ 
km~s$^{-1}$, and a smaller subset of ions including $\ion{O}{i}$, 
 $\ion{Si}{iii}$, $\ion{Ca}{ii}$, $\ion{Fe}{ii}$, and $\ion{Co}{ii}$. 
Of these ions, $\ion{Co}{ii}$ clearly dominates the rich structure observed in the 
wavelength regions that correspond to the $H$ and $K$ passbands.

\begin{figure}[h]
\centering
\includegraphics[width=5.0in]{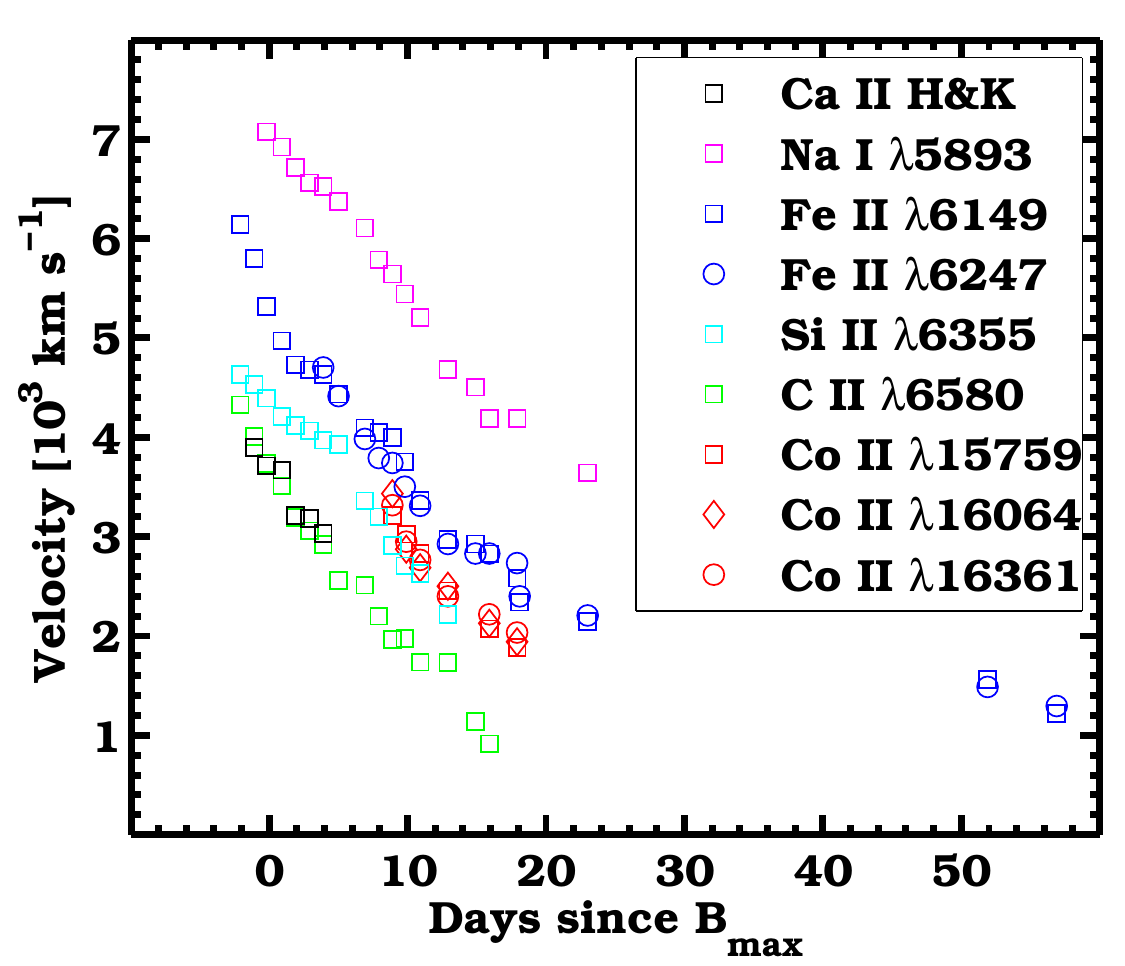}
\caption[]{Velocity evolution of the absorption  minimum of IMEs and Fe-group spectral features that suffer minimal line blending.\label{velocity}}
\end{figure}

Figure~\ref{velocity} exhibits the time evolution of the blue-shifted line velocities 
($v_{exp}$) for a set of ions that suffer minimal to no line blending. 
Specifically, $v_{exp}$ is plotted for IMEs and Fe-group ions located at optical wavelengths including
 $\ion{Ca}{ii}$ $H\&K$,  $\ion{Na}{i}$   $\lambda$5893,   $\ion{Si}{ii}$ $\lambda$6355, $\ion{C}{ii}$ $\lambda$6580, $\ion{Fe}{ii}$ $\lambda$6149\, and $\ion{Fe}{ii}$ $\lambda$6247;  also plotted are $v_{exp}$ values associated with  
 NIR $\ion{Co}{ii}$ features at 15759~\AA,  16064~\AA, and 16361~\AA.
  Around maximum  the IMEs exhibit $v_{exp}$ values that  range from  $\approx$ 4300 km~s$^{-1}$ up to $\approx$ 7100 km~s$^{-1}$, while  $\ion{Fe}{ii}$ $\lambda$6149   
 exhibits $v_{exp} \approx$ 6140 km~s$^{-1}$.
 As the SN evolves, $v_{exp}$ is observed to decrease for all features, with the $\ion{Fe}{ii}$ $\lambda$$\lambda$6149, 6247 features being observed in the 
 $+$57d spectrum to reach $v_{exp}$ $\sim$ 1250 km~s$^{-1}$.
Between $+9$d to $+18$d as the $\ion{Co}{ii}$ lines emerge and dominate the NIR spectrum  their measured  line velocities appear consistent  with 
an averaged value ranging from $\sim$ 3300 to 2000 km~s$^{-1}$, which 
is $\sim$ 800 km~s$^{-1}$ below those inferred   from the optical $\ion{Fe}{ii}$ features.

 \subsection{Spectral Comparison of SN~2010ae to relevant SNe~Iax}

\begin{figure}[h]
\begin{center}$
\begin{array}{cc}
\includegraphics[scale=0.5]{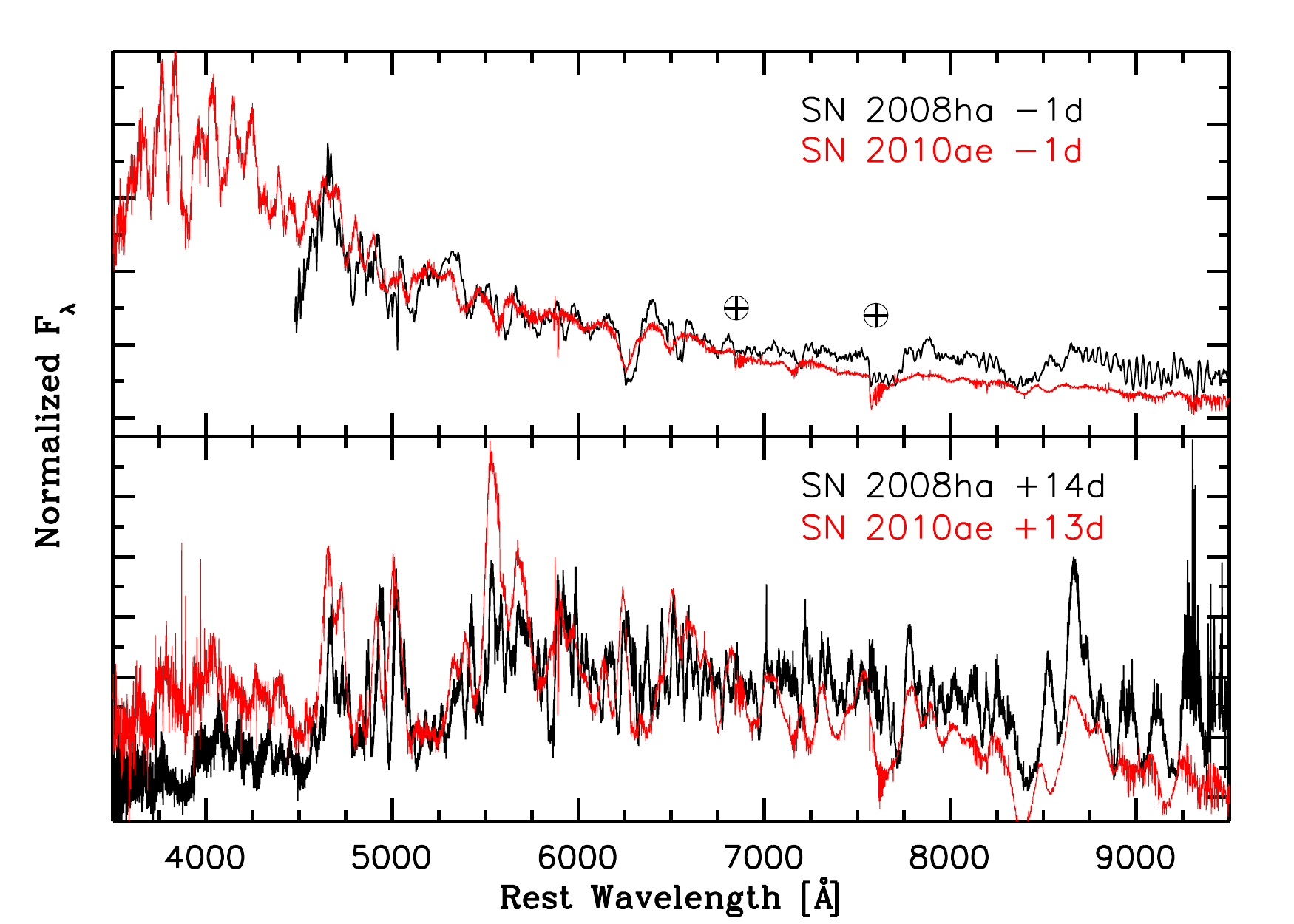} 
\includegraphics[scale=0.5]{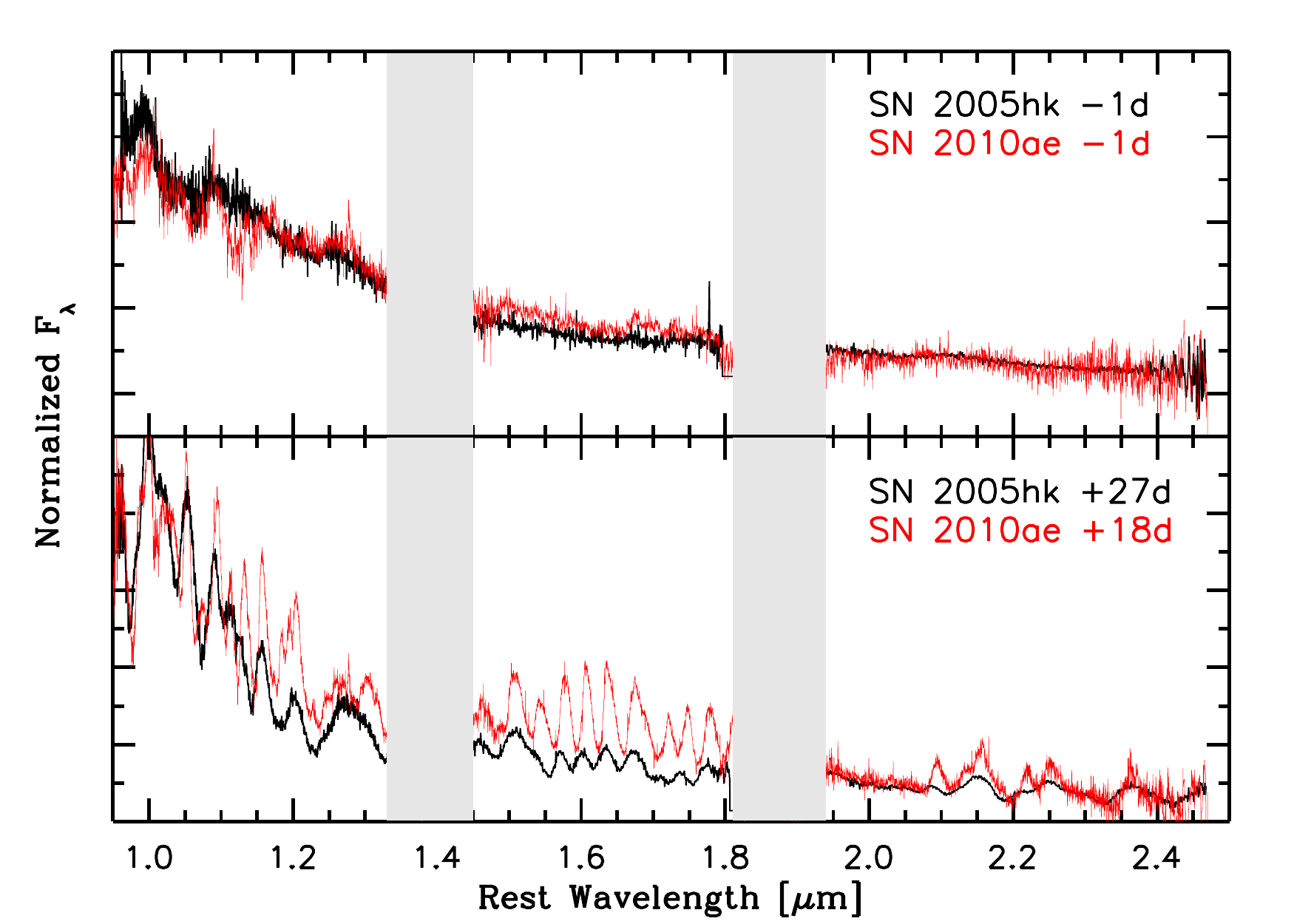} \\
\end{array}$
\end{center}
\caption{ {\em (left)} Comparison between  visual-wavelength spectra of the Type~Iax SNe 2008ha (black) and 2010ae (red)
taken around maximum light and a fortnight later. 
The spectra of SNe~2008ha and 2010ae have been corrected to the rest frame adopting   redshift values $z =$ 0.0047 and $z =$ 0.0037, respectively. 
The spectra of SN~2010ae have also been de-reddened for an $E(B-V)_{tot} = 0.62$ mag. Telluric features are indicated with an Earth symbol.  {\em (right)} Comparison between  NIR spectra of the Type~Iax SNe 2005hk (black) and 2010ae (red)
obtained around maximum light and several weeks later. 
The spectra of SNe~2005hk and 2010ae have been corrected to the rest frame adopting   redshift values $z =$ 0.0130 and $z =$ 0.0037, respectively. 
Vertical  gray bands mask  the most prevalent telluric absorption features.}
\label{speccomp}
\end{figure}

We now proceed to compare spectra of SN~2010ae to other relevant SNe~Iax.
Plotted in the left panel of  Figure~\ref{speccomp}  are the $-1$d and $+$14d visual-wavelength spectra of SN~2010ae compared to similar epoch spectra of SN~2008ha \citep{foley09,foley10a}. Overall the comparison reveals that 
 these objects are spectroscopically similar, particularly at maximum. 
 Close inspection indicates that the main
difference between these two transients is the blue-shifts of the absorption lines, with
 SN~2010ae  exhibiting  higher expansion velocities.
 This results in an  enhancement of  line blending 
  of  many of the low-velocity spectral features that are more clearly resolved in 
   SN~2008ha (see bottom left panel of Figure~\ref{speccomp}). 
   Nevertheless, the most prevalent ions in both objects are in close agreement. 
    
   Turning our attention to redder wavelengths, unfortunately no NIR spectra of SN~2008ha were obtained preventing a one-to-one comparison.
   Indeed NIR spectra of  the spectroscopic defined 
  SNe~Iax   class 
  are nearly nonexistent, with the only published  observations  being  those consisting of a  sequence of five epochs 
  taken of the  bright SN~2005hk  (peak $M_V \sim -18$ mag; \citealt{phillips07}), which were recently presented by \citet{kromer13}.
 Making use of a subset of that unique data set, plotted in the right 
 panel of Figure~\ref{speccomp} is a comparison between NIR 
 spectra of SN~2005hk and  SN~2010ae obtained just prior to maximum light and around three weeks later. 
 Interestingly, although SN~2005hk is more than 3 mag brighter than SN~2010ae at maximum, their NIR spectra are very similar.
 At $-$1d the spectra are characterized by a relatively smooth continuum with no prevalent features red-wards of $\sim$1.2 $\mu$m. 
 This is  consistent with what is observed in other normal SN~Ia, however at much earlier epochs \citep[see e.g.][]{hsiao13}. 
 As revealed in the bottom panel, the $+$27d spectrum of SN~2005hk  exhibits many of the same Fe-group (mostly cobalt) spectral features as in the $+$18d spectrum of SN~2010ae,  though these features become more prominent in SN~2010ae on a shorter timescale.

\subsection{Late-phase spectrum}

\begin{figure}[h]
\centering
\includegraphics[width=7.5in]{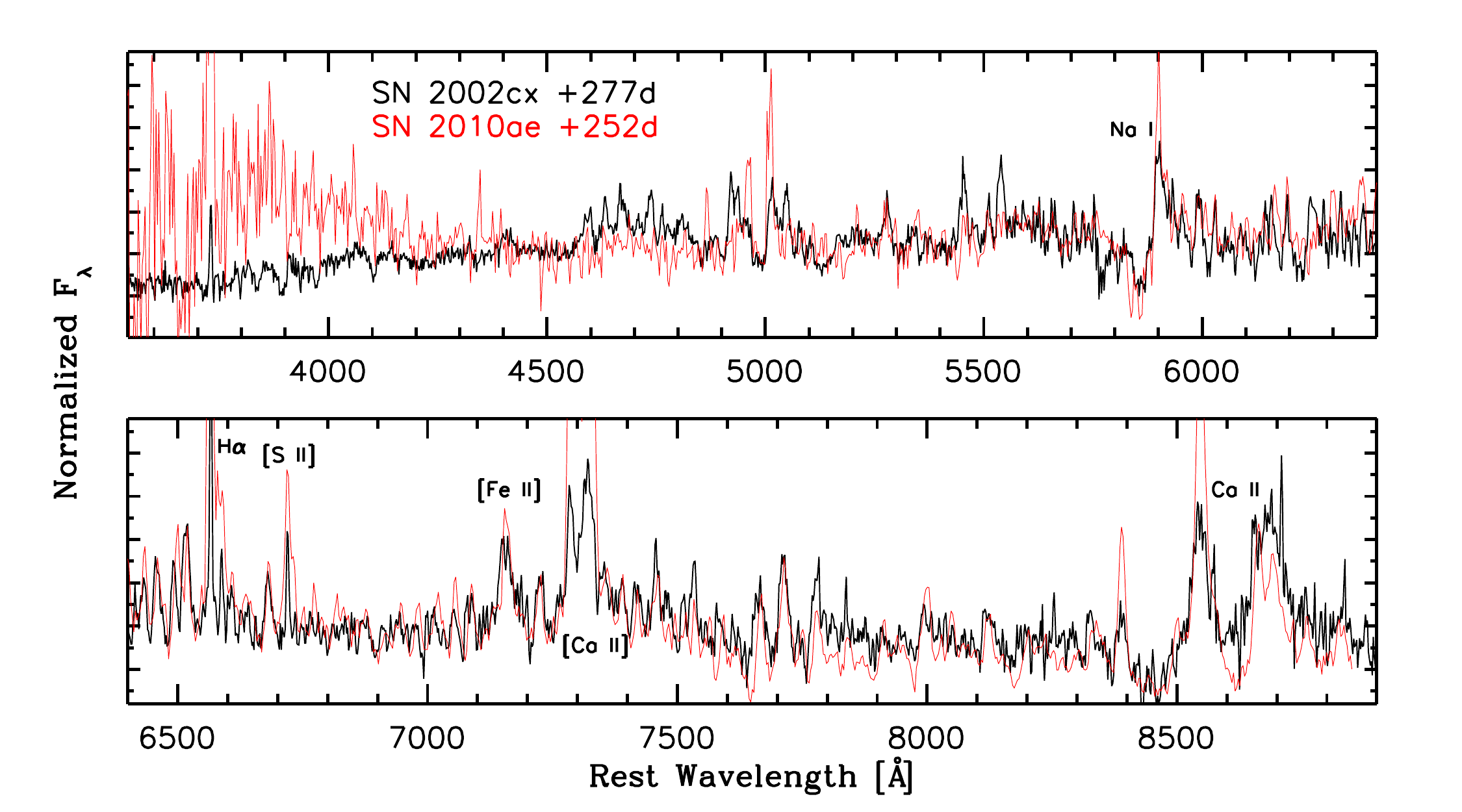}
\caption[]{Expanded view of the late-time ($+$252d) visual-wavelength spectrum of SN~2010ae (red)  compared to a similar epoch spectrum ($+$277d) of  
 the bright Type~Iax SN~2002cx (black)  \citep{jha06}.\label{latespectra}}
\end{figure}

The late-phase  spectrum of SN~2010ae
provides a unique opportunity to examine 
the  nature of an SN~Iax positioned at the faint end of the population. 
As is evident from Figure~\ref{latespectra}, the $+$252d spectrum of SN~2010ae 
bares a striking resemblance to a similarly aged spectrum of the brighter, prototypical 
Type~Iax SN~2002cx.
Like  SN~2002cx, the spectrum of SN~2010ae   exhibits many
low-velocity features associated with both forbidden and P-Cygni permitted \ion{Fe}{ii} lines \citep[see][]{jha06}. 
The majority of cooling, however, appears to occur through the  most prominent emission features being  $[\ion{Ca}{ii}]~\lambda$$\lambda$7291, 7324, and the $\ion{Ca}{ii}$ NIR triplet.
Additional prevalent features include a P-Cygni line located at $\approx$ 5870 \AA\ that is probably attributed to $\ion{Na}{i}~D$  (however $[\ion{Co}{ii}]~\lambda$$\lambda$5890, 5908 cannot be entirely excluded) and forbidden iron lines 
including $[\ion{Fe}{ii}]~\lambda$$\lambda$7155, 7453.

Under close inspection of  Figure~\ref{latespectra}, SN~2010ae does reveal 
 differences with respect to SN~2002cx,
the largest  being between the [\ion{Ca}{ii}] and the \ion{Ca}{ii} NIR triplet emission features, with  both exhibiting different line ratios and being significantly more prominent in SN~2010ae. 
The cause for these differences is probably  related to differences in the 
physical nature of the underlying ejecta.  
Turning to the  analytical work  by \citet{li93},
 we attempt to understand these calcium  line differences through 
 the comparison of the allowed temperature vs. number density relation of the
 underlying ejecta. 
  The adopted relations which were developed under a number of assumptions, including for  example that
 $N_e$ lies within the regime between the critical density of [$\ion{Ca}{ii}$] (i.e. $N_e > 10^{5}$ cm$^{3}$) and  that of the critical density of the \ion{Ca}{ii} NIR triplet (i.e., $N_e < 10^{11}$ / $\tau_{\ion{Ca}{ii}}$ $\sim$ 10$^{8-9}$ cm$^{3}$). 
 In other words, Eq. 21 of \citet{li93}, which uses 
 the ratio ([$\ion{Ca}{ii}~\lambda7291 + \lambda7324]) / \ion{Ca}{ii}~\lambda8542$ as an observational input, is 
 assumed to be roughly valid for $N_e$ values between 
 10$^{5} < N_e < 10^{9}$ cm$^{3}$. 
 Therefore, by assuming that the ratio of the underlying $T_e$ of the ejecta for both objects is at or near unity, and plugging in the appropriate measured calcium ratios, 
 suggests that the $N_e$ of SN~2010ae is approximately a factor 4 smaller than in SN~2002cx. Alternatively, if the values of  $N_e$ for the two objects are fixed to a constant value, this would  suggest that the emitting region of SN~2010ae is $\sim$~2000~K less than that of SN~2002cx. We note that these findings are rather insensitive to the range of 
 reddening values that are adopted in this study.
 From this brief analysis we find that at  similar late epochs, 
 the underlying ejecta of SN~2010ae  is probably less dense and/or cooler
 compared to the corresponding line forming regions of SN~2002cx.
  
\begin{figure}[h]
\centering
\includegraphics[width=7.5in]{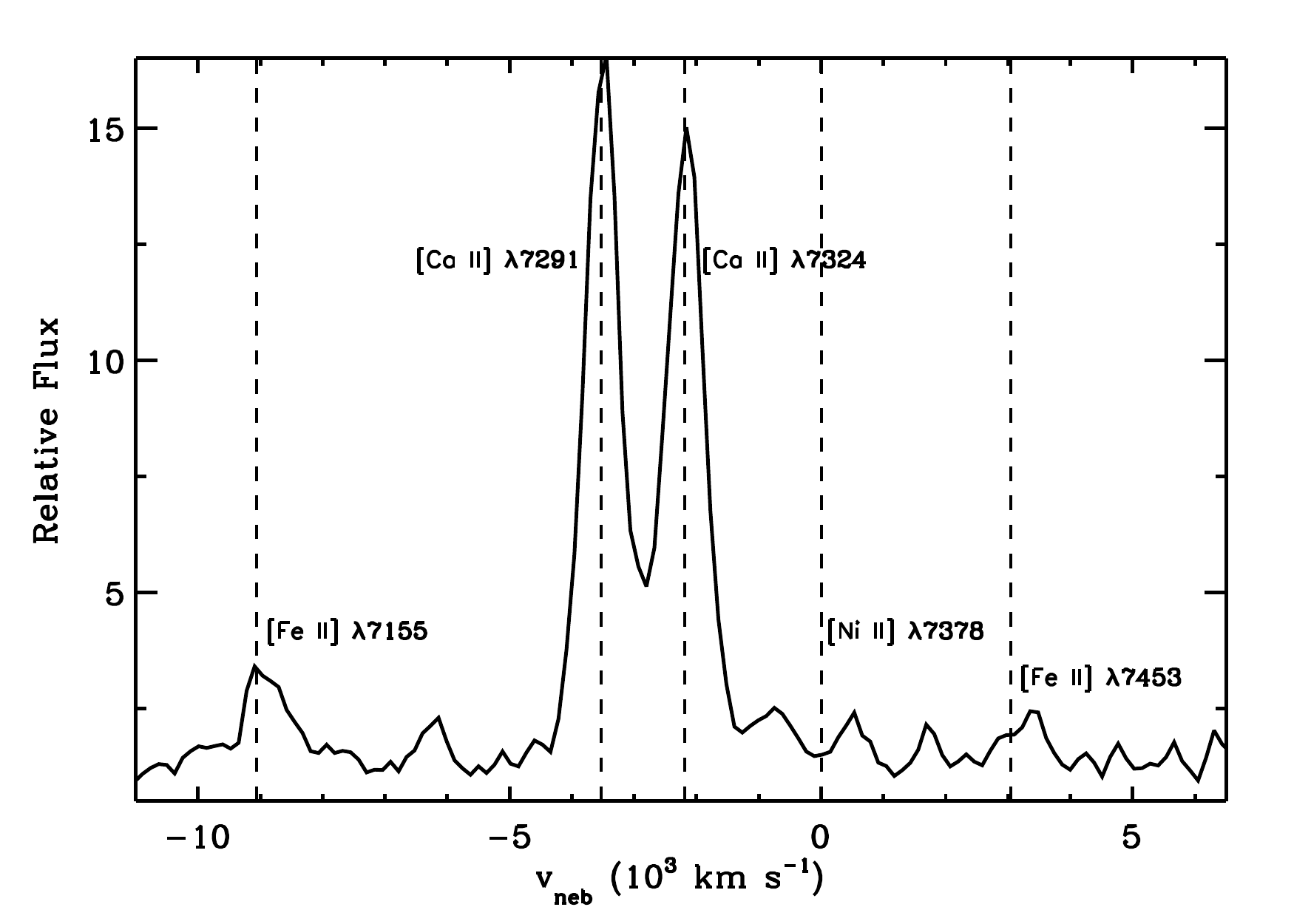}
\caption[]{Late-phase spectrum of SN~2010ae centered on the wavelength region 
around $\sim$ 7300~\AA.\label{lineprofiles}}
\end{figure}

 Plotted in Figure~\ref{lineprofiles} is the late-phase spectrum of SN~2010ae
 zoomed in on the wavelength region around the most prominent $[\ion{Ca}{ii}]$ features, with zero velocity placed at the expected location of $[\ion{Ni}{ii}]$ $\lambda$7378.
  This ion   has been associated with a  prevalent feature in some SNe~Iax \citep[see][their Figure~24]{foley13}, but is   clearly
 not present in SN~2010ae nor in SN~2002cx (see Figure~\ref{latespectra}).
 Also indicated in Figure~\ref{lineprofiles} are the expected locations of $[\ion{Fe}{ii}]$ $\lambda$$\lambda$7155, 7453, both of which have been observed in other SNe~Iax and core-collapse SNe. 
 The width of the $[\ion{Ca}{ii}]$ lines is representative of the majority of emission lines that characterize the late-phase spectrum, which exhibit FWHM (Full-Width-Half-Maximum) velocities ranging between $\sim$~700 to 1000 km~s$^{-1}$, and in addition, show  little indication of significant line shifts and/or large scale asymmetries.

In Figure~\ref{nebcomp} the late-phase spectra of SNe~2002cx and 2010ae are compared to other SNe types at similar epochs. 
The comparison objects include the normal Type~Ia SN~1998bu \citep{silverman13}, 
the under-luminous Type~IIP SN~2008bk, and the Type~Ib SN~2007Y \citep{stritzinger09}. Overall, the comparison indicates that  SNe~Iax 
are in a class of their own. 
As telling as the spectral features identified  in the late-phase spectrum of 
SN~2010ae may or may not be, equally telling  are those features that are not present.
For instance the late-phase spectrum of SN~1998bu shown in Figure~\ref{nebcomp}
contains  broad emission features associated with the blending of 
Fe-group features of $[\ion{Fe}{iii}]$ at $\sim$ 4650 \AA, and 
$[\ion{Fe}{ii}]+[\ion{Fe}{iii}]$ at $\sim$ 5300 \AA. 
As ubiquitous as these features are to thermonuclear SNe, they clearly 
are not evident in the $+$252d spectrum of SN~2010ae, nor for that matter 
in all other SNe~Iax observed at late phases.
Continuing to oxygen, interestingly the spectrum of SN~2010ae shows no evidence for 
$[\ion{O}{i}]$ $\lambda$$\lambda$6300, 6364.
This feature typically dominates the late-phase spectra in the vast majority of core-collapse SNe (see for example SN~2007Y and to lesser extent SN~2008bk in Figure~\ref{nebcomp}), while it is typically not present in late spectra of thermonuclear SNe~Ia
\citep[e.g.][]{blondin12,silverman13}. 
Recently, however,  $[\ion{O}{i}]$  has been observed in the sub-luminous 
Type~Ia SN~2010lp \citep{taubenberger13}, and is expected to be observed in turbulent deflagration explosions of C/O white dwarfs, the possible
 progenitor candidates for  bright 2002cx-like SNe~Iax
\citep[e.g.][]{phillips07,kromer13,fink13}.

\section{Discussion}
\label{sec:discussion}

The  observations presented in this paper confirm
 the existence of another member of a  population of fast evolving, low kinetic energy and low luminosity  objects similar to SN~2008ha that have hitherto  been  missed by transient surveys. To place into context the extreme nature of these faint and fast  objects with other SNe, plotted in Figure~\ref{dm15vslum}
 is a comparison between the peak absolute $B$-band magnitude vs. 
  $\Delta$m$_{15}$ for an extended sample of  bright, normal and faint SNe~Ia observed by the CSP, along with  a handful of other well-observed SNe~Iax. 
 As confirmed from the figure, SN~2008ha is no longer alone at the extreme end of 
 the luminosity vs. decline-rate relation diagram.
 
Our  model fits to the UVOIR light curves of SNe~2008ha and 2010ae 
indicates that they have both produced  a few$\times$10$^{-3}$ M$_{\sun}$ of $^{56}$Ni.
 Although the amount of  $^{56}$Ni is often correlated with effective temperature of the spectra \citep{nugent97,hoeflich96}, here we have a low $^{56}$Ni mass, which leads to a rapidly receding photosphere. In turn, the effective diffusion time is quite low, and the heating of the ejecta is driven by the decay of $^{56}$Ni rather than by $^{56}$Co. 
 As $^{56}$Ni has a much shorter half-life than $^{56}$Co, there is more power available per gram of material \citep{hoeflich93}. This has the net effect of producing  early photospheric spectra that resemble hot and luminous SNe~Ia. 
 
Additional insight within the SNe~Iax class is provided from our extended observations of 
SN~2010ae.
The detailed NIR spectral time series of SN~2010ae has allowed us to 
identify a significant number of Fe-group features that we associate 
predominately with  cobalt. 
Plotted in Figure~\ref{co2} is a comparison of post maximum 
$H$-band NIR spectroscopy of
the normal SN~1999ee \citep{hamuy02}, the sub-luminous 1991bg-like  
SN~1999by \citep{hoeflich02}, and the Type~Iax SNe 
 2005hk \citep{kromer13}, 2008ge, 2010ae, and 2012Z.
 The previously unpublished spectra of SN~2008ge and SN~2012Z were obtained  
 with the VLT equipped with ISAAC (Infrared Spectrometer And Array Camera), and reduced in a standard manner.  
 As discussed earlier in the case of SN~2010ae, a forest of  $\ion{Co}{ii}$ lines 
 dominate this spectral region, and this association evidently  holds true for the more 
  luminous Type~Iax SN~2008ge (peak $M_V \sim -17.4$ mag; \citealt{foley10c}),  SN~2005hk (peak $M_V \sim -18.0$ mag; \citealt{phillips07}), and SN~2012Z (peak $M_V \sim -18.5$ mag; Stritzinger et al., in preparation). 
 
 A cobalt footprint appears to be ubiquitous to this class of objects and provides 
 additional confirmation that the  faint and fast SN~2010ae is indeed spectroscopically similar to the brighter Type~Iax objects like SNe~2002cx and 2005hk. 
 Interestingly, as revealed in the comparison of Figure~\ref{co2},
  the structure and distinctiveness of  the $\ion{Co}{ii}$ forest appears to 
  increase as we go down in the plot. 
  This is probably related to the kinetic energy  of the explosion with the  least energetic  objects suffering less line blending effects. 
  The separation of the features depends on the differential expansion rate of the 
  Fe-group element region combined with the wavelength separation of the multiplets.  
   In normal SNe~Ia where the differential expansion rate is high, these features are 
  blended  leading to the characteristic features observed \citep{wheeler98}. 
  However, at  longer wavelength ($\sim$ 2 $\mu$m) the Fe-group element multiplets are only partially blended  allowing the individual multiplets to be identified. 
 Even with the large differential expansion rate of normal SNe~Ia, 
 the features are reasonable well separated. 
 On the other hand, for sub-luminous 1991bg-like SNe~Ia  the Fe-group elements are  more confined in velocity space, allowing for the identification of individual blended Fe-group multiplets.
 For SN~2010ae the Fe-group elements are even more confined in velocity space, leading to the well-separated features associated with the individual multiplets. 
      
\begin{figure}[h]
\centering
\includegraphics[width=5.5in]{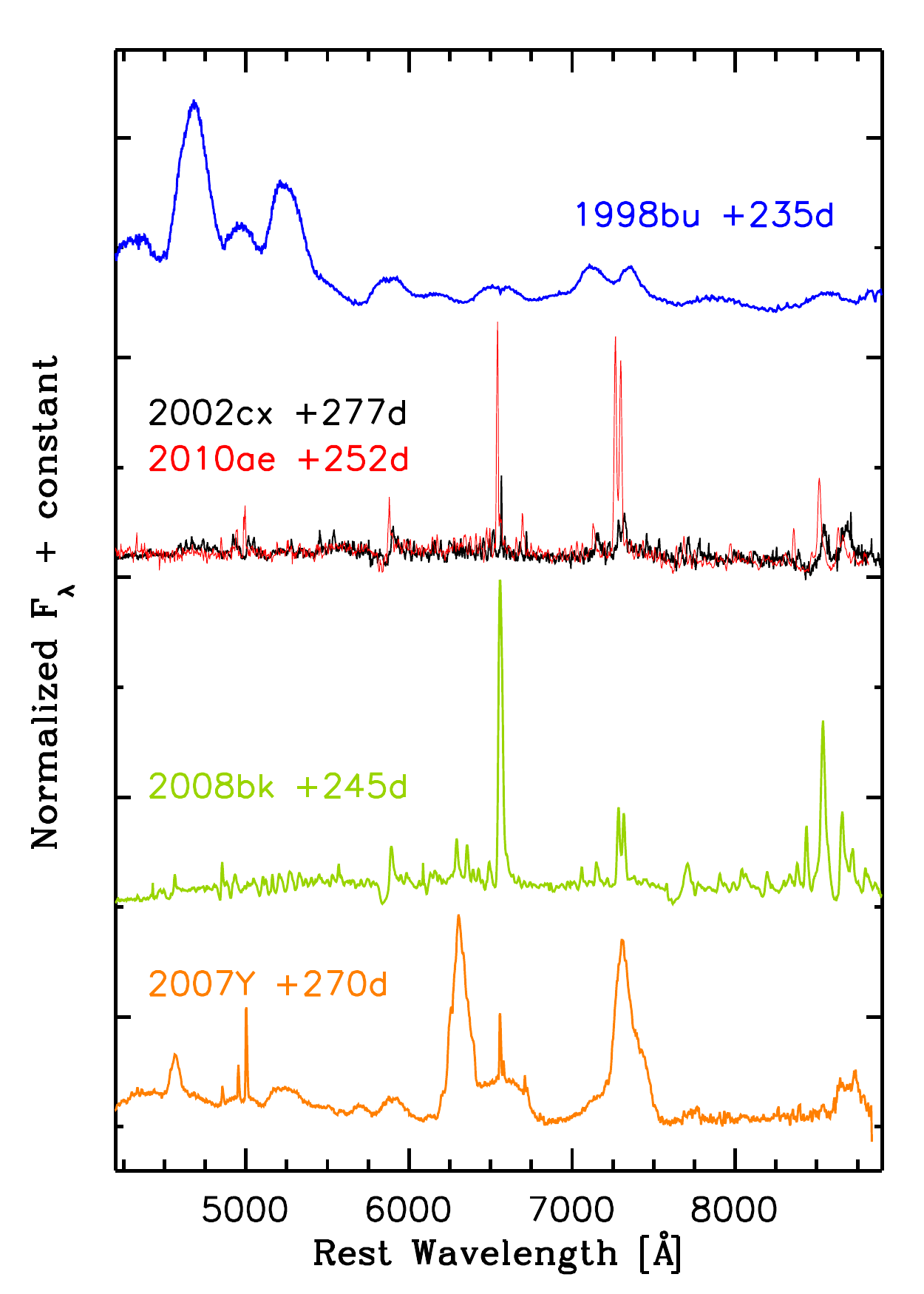}
\caption[]{Comparison of the late-phase spectra of the Type~Iax SNe~2002cx and 
2010ae to similar epoch spectra of the normal Type~Ia SN~1998bu \citep{silverman13}, the under-luminous  Type~IIP SN~2008bk (CSP, unpublished), and the Type~Ib SN~2007Y \citep{stritzinger09}. Each spectrum is labeled with respect to $T(B_{max})$, except for SN~2008bk where its epoch is with respect to the date of discovery. \label{nebcomp}}
\end{figure}

\begin{figure}[h]
 \centering
\includegraphics[width=7.5in]{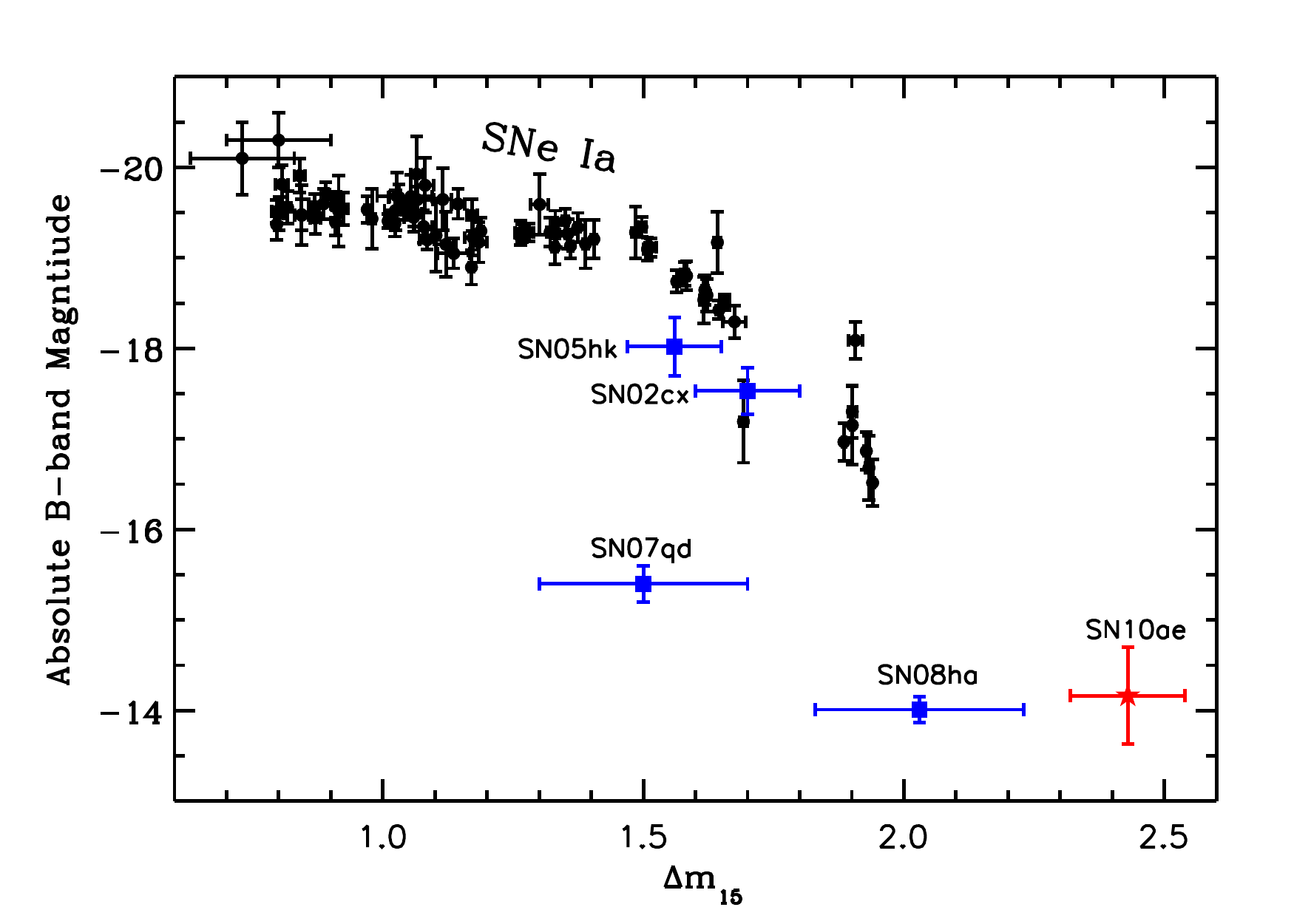}
\caption[]{Peak absolute $B$-band magnitude vs. $\Delta$m$_{15}$ for a sample of  
CSP SNe~Ia  (black  dots), several  SNe~Iax (blue  squares), and SN~2010ae (red star). The SNe~Iax plotted  from  brightest to faintest  are SN~2005hk \citep{phillips07}, SN~2002cx \citep{li03,phillips07}, SN~2007qd  \citep{mcclelland10}, and SN~2008ha.
\label{dm15vslum}}
\end{figure}

\begin{figure}[h]
\centering
\includegraphics[width=6.in]{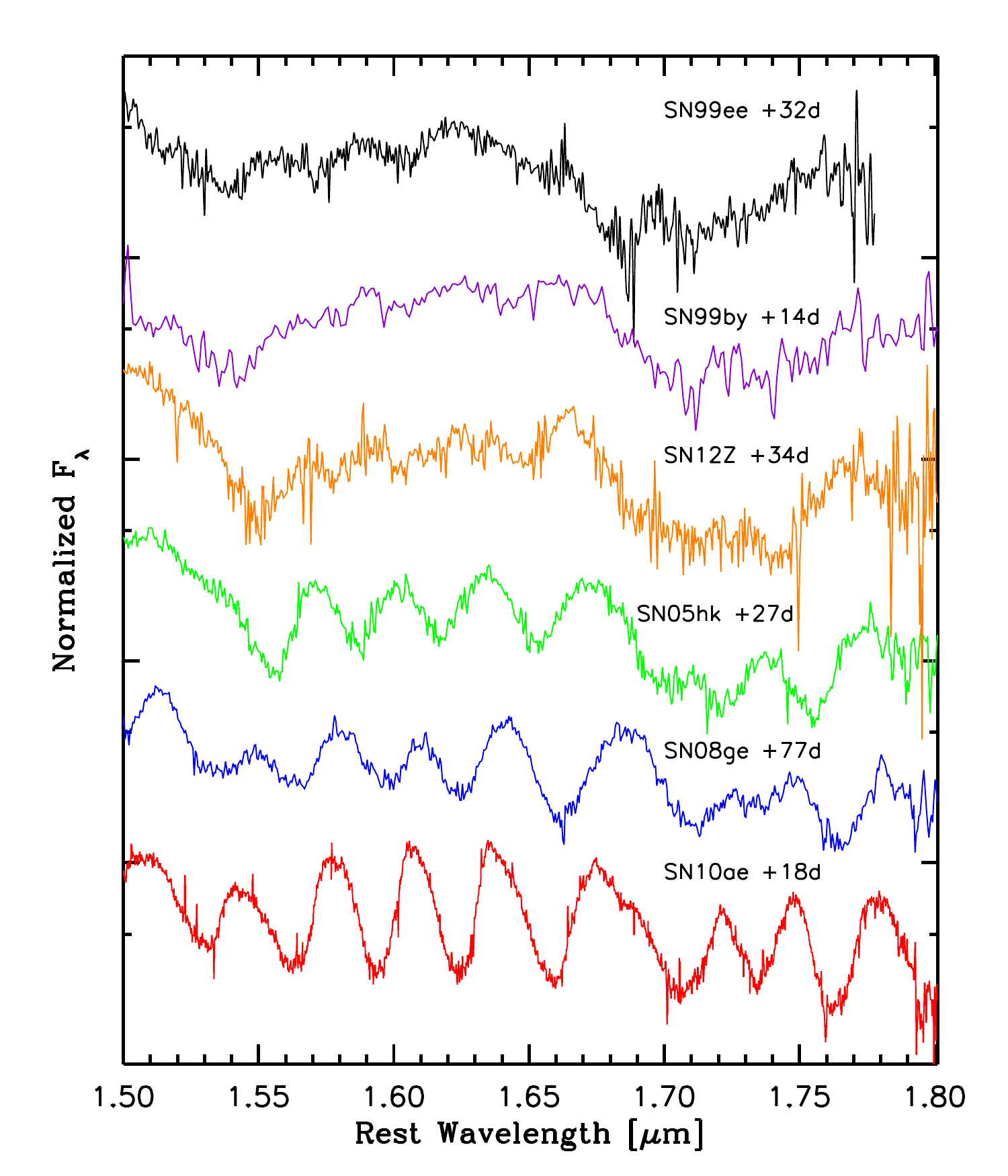}
\caption[]{$H$-band NIR spectroscopy of the normal SN~1999ee \citep{hamuy02} and
the sub-luminous Type~Ia SN~1999by \citep{hoeflich02}, 
compared to  the Type~Iax SNe 2012Z, 2005hk \citep{kromer13}, 2008ge, and 2010ae.
The SNe~Iax spectra are ordered (top to bottom)  from brightest to faintest, and the epoch 
with respect to maximum light is  indicated for  each object.
$\ion{Co}{ii}$ features appear to be a ubiquitous signature of SNe~Iax class, showing increased prominence  in the lowest velocity objects.\label{co2}}
\end{figure}

  Although the Type~Iax spectral class exhibits some homogeneity there
  are a number of observational signatures that demonstrate 
   variety, such as that revealed through the comparison of  late-phase spectra.  
We now turn to the presence of $[\ion{Fe}{ii}]$ $\lambda$7155 and  
$[\ion{Ni}{ii}]$ $\lambda$7378 emission lines, both of which   are  thought to be formed in the ashes of a deflagration flame \citep[e.g.][]{maeda10}, and thereby provide 
some clues to the burning physics. 
Both features are clearly conspicuous  in a handful of  SNe~Iax  \citep{foley13}. However, through examination of the late-phase spectra of the bright SN~2002cx and now the faint  SN~2010ae,  we find little to no evidence for  $[\ion{Ni}{ii}]$ $\lambda$7378 emission, while 
  $[\ion{Fe}{ii}]$ $\lambda$7155 is discernible, albeit not at the strength seen in other objects, e.g. SNe 2005P and 2008ge \citep[see][]{foley10c,foley13}. 
A lack of forbidden lines implies high densities consistent with a fast receding 
 photosphere into the high density regions. 
   Additionally, our analysis of the dominant cooling calcium features provides an indication that there are clear differences in the physical conditions of the underlying ejecta. Additional late-phase observations of other low- and high-luminosity SNe~Iax are required to determine how cooling is dependent on the physical parameters of the ejecta. 
  Summarizing, the late-phase spectrum of SN~2010ae resembles SN~2002cx;  the main differences  observed are the strength and line ratios of the calcium features. 
 Finally, we note that all late-phase spectra obtained to date of spectroscopically classified SNe~Iax  suggest a relative dense ejecta at low velocity \citep{jha06,foley13,mccully13},
  and are substantially different from the spectra of similar aged thermonuclear and core-collapse SNe (see Figure~\ref{nebcomp}).   
 
 Initial  studies of the bright and energetic Type~Iax SNe~2002cx and 2005hk
 suggested that their observational properties were most consistent with 
 3-D deflagration models \citep[e.g.][]{branch04,phillips07}.
Recently, a suite of modeled calculations of  Chandrasekhar mass
 carbon-oxygen white dwarf models that burn via a turbulent deflagration flame 
 and leave a bound remnant has been published \citep{kromer13,fink13,jordan12}.
 These models provide a promising avenue to explain the range of  SN~Iax luminosities and explosion energies, however, they  appear to produce $^{56}$Ni masses that are 
 roughly an order of magnitude larger than the values inferred from the observations of
 SNe~2008ha and 2010ae.
For these extremely low-luminosity, low-ejecta mass objects   
 their origins may be more attuned to the partial thermonuclear incineration of helium-accreting {\em sub}-Chandrasekhar mass carbon-oxygen white dwarfs that burn via a deflagration flame, and leave a bound remnant \citep[see e.g.][]{foley13}.
  Unfortunately, the viability of this progenitor  path remains  at best conjecture,
  and additional modeling efforts including  full radiative transfer calculations are  required to determine if SN can be simulated that create the low  $^{56}$Ni masses inferred for the faintest events.

Past efforts to model sub-luminous SNe~Ia consist of edge-lit explosions where 
a helium layer is accreted onto either a sub-Chandrasekhar or a Chandrasekhar mass carbon-oxygen white dwarf, and subsequently
initiates a detonation \citep{livne90,woosley94,livne95} or a 
double detonation explosion \citep{fink10,sim12,townsley12}.
Within this realm, even less robust explosions could occur if the edge-lit explosion does not 
successfully unbind the white dwarf \citep{bildsten07,shen10,wang13a,wang13b}. 
However, it is unclear if any variant within this extended family of models can successfully reproduce all of the most important  observables of the least luminous SNe~Iax, including low velocities, low $^{56}$Ni mass, and the synthesis of IMEs.
  
  Apart from a white dwarf origin others have explored the possibility 
 of fallback models \citep{valenti08,moriya10}, as well as electron-capture 
 SNe from super-AGB (Asymptotic Giant Branch) progenitors \citep{pumo10}, but these avenues appear less likely given the  range of observational properties displayed by the SN~Iax class.

 Future efforts to model this unique class of stellar explosions will have to consider
 the broad range of observational details touched on in this study
  before a consensus can be reached about whether nature prefers a unique or a multiple progenitor path to produce the range of 
  objects that fall under the SNe~Iax designation.  
  In a forthcoming paper we will present model calculations for SN~2010ae.

\begin{acknowledgements}
Special thanks to the referee who provided a very useful report that improved the
quality of this publication.
The authors are grateful to R. J. Foley, V. Stanishev and I. R. Seitenzahl for stimulating discussions and 
providing access to published spectra, as well as  to CSP observers L. Boldt, S. Castellon, F. Salgado  and  W. Krzeminski and staff Astronomers at the Gemini South and ESO VLT observatories for performing observations.
M. D. S. and C. C. gratefully acknowledge generous support provided by the Danish Agency for Science and Technology and Innovation  
realized through a Sapere Aude Level 2 grant.
M. D. S., S. V. and F. T. acknowledge funding provided by the Instrument Center for Danish Astrophysics (IDA).
M. D.~S., K. M. and G. F. acknowledge support by World Premier International Research Center Initiative, MEXT, Japan. 
G. L. is supported by the Swedish Research Council through grant No. 623-2011-7117.
A. P. and S.B. are partially supported by the PRIN-INAF 2011 with the project ``Transient Universe: from ESO Large to PESSTO".
G. P. acknowledges funding provided by the Proyecto FONDECYT 11090421.
M. H. and G. P. acknowledge support provided by the Millennium Center for Supernova Science through grant P10-064-F (funded by Programa Iniciativa Cientifica Milenio del Ministerio de Economia, Fomento y Turismo de Chile).
This material is also based upon work supported by NSF under 
grants AST--0306969, AST--0607438 and AST--1008343. 
This research has made use of the NASA/IPAC Extragalactic Database (NED), which is operated by the Jet Propulsion Laboratory, California Institute of Technology, under contract with the National Aeronautics and Space Administration; as well as 
resources from the National Energy Research Scientific Computing Center (NERSC), which is supported by the Office of Science of the U.S. Department of Energy under Contract No. DE-AC02-05CH11231.

\end{acknowledgements}

\bibliographystyle{aa}

\clearpage

\clearpage

\clearpage

\clearpage

\clearpage

\clearpage

\clearpage

\clearpage

\clearpage

\clearpage

\clearpage

\clearpage

\clearpage

\clearpage

\clearpage

\clearpage

\clearpage
\input{SN10ae_locseq.tex}

\clearpage 
\input{SN10ae_optphot.tex}

\clearpage
  \input{SN10ae_nirphot.tex}

\clearpage
\input{specjournal.tex}

\clearpage
\input{lcparameters.tex}

\clearpage
\appendix
\section{Supernova 2008ha}

SN~2008ha was discovered in the irregular  galaxy UGC~12682  on 2008 November 7.17 UT   through the course of the Puckett Observatory Supernova Search  \citep{puckett10}.
With  J2000.0  coordinates of 
$\alpha$ $=$ 23$^{\rm h}$34$^{\rm m}$52\fs69 and 
$\delta = +$18$^{\circ}$13$\arcmin$35\farcs4, the SN was
positioned approximately 12\arcsec\ West and 0\farcs5 South from the 
center of its host. 
According to NED, the \citet{schlafly11} recalibration of the \citet{schlegel98} dust maps provides a visual extinction value 
$A_{V} =$ 0.21 mag. This value is adopted in our analysis, and is slightly lower than  the Schlegel dust maps value (adopted by \citealt{valenti09} and \citealt{foley09}) of $A_{V} =$ 0.25 mag.
To set the absolute flux scale of SN~2008ha    
the NED Virgo Infall corrected redshift distance of 1553$\pm$17 km~s$^{-1}$ is adopted, which for an $H_{\circ} = 73\pm$5 km~s$^{-1}$~Mpc$^{-1}$, corresponds to 
21.3$\pm$1.5 Mpc    or $\mu = 31.64\pm$0.15 mag.

Optical ($u'g'r'i'$) and NIR ($YJH$) imaging of SN~2008ha was obtained at the Las Campanas Observatory using facilities available to the {\em Carnegie Supernova Project}. 
Depending on the particular filter as many as 18 epochs of optical imaging was obtained with the Swope telescope covering the flux evolution from  approximately $-$1d to $+$28d  relative to $T(B)_{max}$. 
Our NIR follow up was considerably more sparse consisting of five epochs of imaging with the Swope ($+$RetroCam) ranging from $+$0.2d to $+$18.1d relative to T(B)$_{max}$.
 The data were processed in the standard manner, including template subtraction, following the methods described in \citet{contreras10} and \citet{stritzinger11}. 

Differential PSF photometry of the SN was computed from template subtracted 
science images relative to a local sequence of stars calibrated over the course of multiple photometric nights with respect to the \citet{landolt92}, \citet{persson98}, and  \citet{smith02} standard fields. 
Coordinates  and final magnitudes of the local sequence are provided in 
Table~\ref{SN08halocalsequence}. Definitive optical and NIR photometry of SN~2008ha in the standard photometric systems are given in Table~\ref{SN08ha_optphot} and Table~\ref{SN08ha_nirphot}, respectively.

\clearpage 
\input{SN08ha_locseq.tex}

\clearpage
\input{SN08ha_optphot.tex}

\clearpage
\input{SN08ha_nirphot.tex}

\end{document}

%% file: SN10ae_locseq.tex
\begin{deluxetable} {cllcccccccccc}
\rotate
\tabletypesize{\tiny}
\tablecolumns{13}
\tablewidth{0pt}
\tablecaption{Optical and NIR photometry of the local sequence of SN~2010ae in the standard system.\label{localsequence}}
\tablehead{
\colhead{}     &
\colhead{}     &
\colhead{}     &
\colhead{$g'$} &
\colhead{$r'$} &
\colhead{$i'$} &
\colhead{$z'$} &
\colhead{$B$}  &
\colhead{$V$}  &
\colhead{$R$}  &
\colhead{$Y$}  &
\colhead{$J$}  &
\colhead{$H$} \\
\colhead{STAR}             &
\colhead{$\alpha~(2000)$}  &
\colhead{$\delta~(2000)$}  &
\colhead{(mag)}            &
\colhead{(mag)}            &
\colhead{(mag)}            &
\colhead{(mag)}            &
\colhead{(mag)}            &
\colhead{(mag)}            &
\colhead{(mag)}            &
\colhead{(mag)}            &
\colhead{(mag)}            &
\colhead{(mag)}}
\startdata
01 & 07$^{\rm h}$15$^{\rm m}$50$\fs$15 & $-$57$^{\circ}$17$\arcmin$25$\farcs$12  & 12.120(026) &  11.800(026) &  11.692(037) &  11.664(019) &  12.430(033) &  11.941(022) &  11.622(016) &  $\cdots$ &   $\cdots$ &  $\cdots$ \\
02 & 07$^{\rm h}$15$^{\rm m}$40$\fs$13 & $-$57$^{\circ}$20$\arcmin$55$\farcs$86  & 14.282(019) &  13.535(020) &  13.279(034) &  13.148(026) &  14.772(018) &  13.862(019) &  13.307(019) &  12.480(009) &   12.174(011) &   11.726(008) \\
03 & 07$^{\rm h}$16$^{\rm m}$14$\fs$17 & $-$57$^{\circ}$20$\arcmin$46$\farcs$18  & 15.269(018) &  14.415(017) &  14.087(033) &  13.901(016) &  15.863(022) &  14.801(021) &  14.174(030) &  13.206(008) &   12.844(010) &   12.322(009) \\
04 & 07$^{\rm h}$16$^{\rm m}$08$\fs$86 & $-$57$^{\circ}$21$\arcmin$49$\farcs$82  & 14.563(026) &  14.184(017) &  14.045(037) &  13.996(012) &  14.891(023) &  14.351(013) &  13.998(012) &  13.405(009) &   13.189(013) &   12.925(008) \\
05 & 07$^{\rm h}$16$^{\rm m}$15$\fs$74 & $-$57$^{\circ}$23$\arcmin$03$\farcs$62  & 15.463(019) &  14.707(027) &  14.369(075) &  14.179(048) &  16.003(024) &  15.055(026) &  14.493(022) &  13.550(009) &   13.190(011) &   12.680(009) \\
06 & 07$^{\rm h}$15$^{\rm m}$48$\fs$12 & $-$57$^{\circ}$19$\arcmin$14$\farcs$41  & 15.807(025) &  15.172(019) &  14.944(027) &  14.830(048) &  16.249(014) &  15.446(025) &  14.960(021) &  14.193(009) &   13.901(014) &   13.491(009) \\
07 & 07$^{\rm h}$16$^{\rm m}$01$\fs$29 & $-$57$^{\circ}$23$\arcmin$16$\farcs$08  & 17.393(765) &  16.366(034) &  15.470(050) &  14.989(053) &  18.617(069) &  16.999(041) &  16.024(011) &  14.212(009) &   13.765(010) &   13.113(009) \\
08 & 07$^{\rm h}$16$^{\rm m}$16$\fs$69 & $-$57$^{\circ}$24$\arcmin$28$\farcs$40  & 16.732(504) &  15.836(014) &  15.561(034) &  15.382(020) &  17.035(029) &  16.160(034) &  15.592(013) &  14.777(014) &   14.437(020) &   13.964(012) \\
09 & 07$^{\rm h}$15$^{\rm m}$57$\fs$29 & $-$57$^{\circ}$21$\arcmin$09$\farcs$22  & 16.780(109) &  15.798(018) &  15.335(041) &  15.118(062) &  17.435(029) &  16.262(026) &  15.509(029) &  14.351(008) &   13.949(012) &   13.355(008) \\
10 & 07$^{\rm h}$15$^{\rm m}$47$\fs$91 & $-$57$^{\circ}$25$\arcmin$12$\farcs$47  & 17.984(044) &  16.724(021) &  16.039(070) &  15.750(066) &  12.423(010) &  17.347(028) &  16.417(028) &  14.915(020) &   14.483(031) &   13.821(012) \\
11 & 07$^{\rm h}$15$^{\rm m}$54$\fs$75 & $-$57$^{\circ}$25$\arcmin$03$\farcs$14  & 16.710(019) &  16.076(031) &  15.815(037) &  15.681(048) &  17.207(038) &  16.362(015) &  15.864(018) &  14.993(020) &   14.690(019) &   14.258(014) \\
12 & 07$^{\rm h}$15$^{\rm m}$50$\fs$86 & $-$57$^{\circ}$23$\arcmin$07$\farcs$30  & 15.838(027) &  15.376(016) &  15.207(031) &  15.111(030) &  16.202(017) &  15.572(022) &  15.187(019) &  14.518(009) &   14.273(011) &   13.966(008) \\
13 & 07$^{\rm h}$15$^{\rm m}$49$\fs$86 & $-$57$^{\circ}$22$\arcmin$53$\farcs$26  & 16.241(025) &  15.669(020) &  15.451(035) &  15.338(033) &  16.653(019) &  15.922(024) &  15.472(016) &  14.715(009) &   14.436(012) &   14.085(009) \\
14 & 07$^{\rm h}$15$^{\rm m}$52$\fs$66 & $-$57$^{\circ}$19$\arcmin$26$\farcs$54  & 17.420(023) &  16.330(033) &  15.884(043) &  15.649(077) &  18.015(024) &  16.835(031) &  16.045(019) &  14.928(008) &   14.531(011) &   13.925(009) \\
15 & 07$^{\rm h}$15$^{\rm m}$58$\fs$59 & $-$57$^{\circ}$22$\arcmin$37$\farcs$78  & 20.367(094) &  18.862(034) &  17.031(061) &  16.126(054) &  $\cdots$    &  $\cdots$    &  18.276(074) &  15.088(008) &   14.538(015) &   13.996(008) \\
16 & 07$^{\rm h}$15$^{\rm m}$51$\fs$49 & $-$57$^{\circ}$22$\arcmin$24$\farcs$96  & 18.009(013) &  16.801(036) &  16.295(018) &  15.981(104) &  18.647(075) &  17.385(019) &  16.515(017) &  15.207(008) &   14.780(009) &   14.126(009) \\
17 & 07$^{\rm h}$16$^{\rm m}$07$\fs$36 & $-$57$^{\circ}$19$\arcmin$32$\farcs$66  & 17.615(013) &  16.851(113) &  16.468(057) &  16.333(067) &  18.145(022) &  17.181(036) &  16.569(027) &  15.600(015) &   15.244(014) &   14.745(009) \\
18 & 07$^{\rm h}$15$^{\rm m}$46$\fs$31 & $-$57$^{\circ}$19$\arcmin$55$\farcs$42  & 16.934(025) &  16.470(039) &  16.285(046) &  16.220(016) &  17.240(022) &  16.683(015) &  16.291(016) &  15.634(009) &   15.408(013) &   15.060(011) \\
19 & 07$^{\rm h}$15$^{\rm m}$47$\fs$01 & $-$57$^{\circ}$20$\arcmin$28$\farcs$25  & 19.130(036) &  17.863(024) &  17.288(069) &  16.901(046) &  19.672(107) &  18.370(048) &  17.576(062) &  16.131(016) &   15.713(013) &   15.026(017) \\
20 & 07$^{\rm h}$16$^{\rm m}$12$\fs$81 & $-$57$^{\circ}$17$\arcmin$26$\farcs$92  & 17.032(026) &  16.487(040) &  16.313(052) &  16.189(054) &  17.453(045) &  $\cdots$    &  $\cdots$    &  15.628(012) &   15.375(017) &   15.049(024) \\
21 & 07$^{\rm h}$16$^{\rm m}$01$\fs$68 & $-$57$^{\circ}$19$\arcmin$14$\farcs$02  & 18.072(028) &  17.565(016) &  17.383(040) &  $\cdots$    &  18.391(046) &  17.762(046) &  17.395(025) &  16.659(016) &   16.429(028) &   16.153(030) \\
22 & 07$^{\rm h}$16$^{\rm m}$09$\fs$04 & $-$57$^{\circ}$19$\arcmin$52$\farcs$21  & 20.440(142) &  19.058(054) &  18.180(045) &  $\cdots$    &  $\cdots$    &  19.700(114) &  18.795(112) &  16.811(018) &   16.341(026) &   15.768(019) \\
23 & 07$^{\rm h}$16$^{\rm m}$20$\fs$30 & $-$57$^{\circ}$21$\arcmin$58$\farcs$14  & $\cdots$    &  $\cdots$    &  $\cdots$    &  $\cdots$    &  $\cdots$    &  $\cdots$    &  $\cdots$    &  12.261(008) &   11.869(016) &   11.304(015) \\
24 & 07$^{\rm h}$16$^{\rm m}$14$\fs$32 & $-$57$^{\circ}$21$\arcmin$41$\farcs$80  & $\cdots$    &  $\cdots$    &  $\cdots$    &  $\cdots$    &  $\cdots$    &  $\cdots$    &  $\cdots$    &  13.611(008) &   13.376(009) &   13.075(009) \\
25 & 07$^{\rm h}$16$^{\rm m}$30$\fs$09 & $-$57$^{\circ}$24$\arcmin$21$\farcs$67  & $\cdots$    &  $\cdots$    &  $\cdots$    &  $\cdots$    &  $\cdots$    &  $\cdots$    &  $\cdots$    &  14.252(012) &   13.976(036) &   13.641(011) \\
26 & 07$^{\rm h}$16$^{\rm m}$09$\fs$64 & $-$57$^{\circ}$25$\arcmin$51$\farcs$38  & $\cdots$    &  $\cdots$    &  $\cdots$    &  $\cdots$    &  $\cdots$    &  $\cdots$    &  $\cdots$    &  14.431(014) &   14.031(020) &   13.540(014) \\
27 & 07$^{\rm h}$15$^{\rm m}$49$\fs$86 & $-$57$^{\circ}$25$\arcmin$39$\farcs$29  & $\cdots$    &  $\cdots$    &  $\cdots$    &  $\cdots$    &  $\cdots$    &  $\cdots$    &  $\cdots$    &  15.202(012) &   14.899(014) &   14.501(014) \\
28 & 07$^{\rm h}$15$^{\rm m}$48$\fs$87 & $-$57$^{\circ}$15$\arcmin$59$\farcs$65  & $\cdots$    &  $\cdots$    &  $\cdots$    &  $\cdots$    &  $\cdots$    &  $\cdots$    &  $\cdots$    &  14.006(012) &   13.621(020) &   13.033(014) \\
29 & 07$^{\rm h}$16$^{\rm m}$16$\fs$80 & $-$57$^{\circ}$25$\arcmin$15$\farcs$24  & $\cdots$    &  $\cdots$    &  $\cdots$    &  $\cdots$    &  $\cdots$    &  $\cdots$    &  $\cdots$    &  15.653(012) &   15.293(032) &   14.789(020) \\
30 & 07$^{\rm h}$16$^{\rm m}$24$\fs$49 & $-$57$^{\circ}$20$\arcmin$2.$\farcs$96  & $\cdots$    &  $\cdots$    &  $\cdots$    &  $\cdots$    &  $\cdots$    &  $\cdots$    &  $\cdots$    &  15.451(011) &   15.142(016) &   14.745(017) \\
31 & 07$^{\rm h}$16$^{\rm m}$19$\fs$23 & $-$57$^{\circ}$18$\arcmin$47$\farcs$16  & $\cdots$    &  $\cdots$    &  $\cdots$    &  $\cdots$    &  $\cdots$    &  $\cdots$    &  $\cdots$    &  15.609(014) &   15.196(018) &   14.601(019) \\
32 & 07$^{\rm h}$16$^{\rm m}$07$\fs$36 & $-$57$^{\circ}$19$\arcmin$32$\farcs$66  & $\cdots$    &  $\cdots$    &  $\cdots$    &  $\cdots$    &  $\cdots$    &  $\cdots$    &  $\cdots$    &  15.600(015) &   15.244(014) &   14.745(009) \\
33 & 07$^{\rm h}$15$^{\rm m}$46$\fs$31 & $-$57$^{\circ}$19$\arcmin$55$\farcs$42  & $\cdots$    &  $\cdots$    &  $\cdots$    &  $\cdots$    &  $\cdots$    &  $\cdots$    &  $\cdots$    &  15.634(009) &   15.408(013) &   15.060(011) \\
34 & 07$^{\rm h}$16$^{\rm m}$25$\fs$39 & $-$57$^{\circ}$25$\arcmin$49$\farcs$84  & $\cdots$    &  $\cdots$    &  $\cdots$    &  $\cdots$    &  $\cdots$    &  $\cdots$    &  $\cdots$    &  16.201(014) &   15.934(026) &   15.553(048) \\
35 & 07$^{\rm h}$16$^{\rm m}$26$\fs$31 & $-$57$^{\circ}$18$\arcmin$52$\farcs$20  & $\cdots$    &  $\cdots$    &  $\cdots$    &  $\cdots$    &  $\cdots$    &  $\cdots$    &  $\cdots$    &  15.750(010) &   15.459(028) &   15.039(026) \\
36 & 07$^{\rm h}$16$^{\rm m}$18$\fs$76 & $-$57$^{\circ}$16$\arcmin$19$\farcs$38  & $\cdots$    &  $\cdots$    &  $\cdots$    &  $\cdots$    &  $\cdots$    &  $\cdots$    &  $\cdots$    &  14.954(012) &   14.653(024) &   14.216(017) \\
37 & 07$^{\rm h}$16$^{\rm m}$02$\fs$61 & $-$57$^{\circ}$15$\arcmin$48$\farcs$38  & $\cdots$    &  $\cdots$    &  $\cdots$    &  $\cdots$    &  $\cdots$    &  $\cdots$    &  $\cdots$    &  15.060(015) &   14.703(020) &   14.235(020) \\
38 & 07$^{\rm h}$16$^{\rm m}$26$\fs$50 & $-$57$^{\circ}$16$\arcmin$35$\farcs$62  & $\cdots$    &  $\cdots$    &  $\cdots$    &  $\cdots$    &  $\cdots$    &  $\cdots$    &  $\cdots$    &  15.323(012) &   15.030(014) &   14.632(014) \\
39 & 07$^{\rm h}$16$^{\rm m}$01$\fs$68 & $-$57$^{\circ}$19$\arcmin$14$\farcs$02  & $\cdots$    &  $\cdots$    &  $\cdots$    &  $\cdots$    &  $\cdots$    &  $\cdots$    &  $\cdots$    &  16.659(016) &   16.429(028) &   16.153(030) \\
\enddata
\tablecomments{Uncertainties given in parentheses in thousandths of a
  magnitude correspond to an rms of the magnitudes obtained on
  photometric nights.}
\end{deluxetable}

%% file: SN10ae_optphot.tex
\clearpage
\begin{deluxetable}{cccccccc}
\tablewidth{0pt}
\tabletypesize{\scriptsize}
\tablecaption{PROMPT optical photometry of SN~2010ae in the standard system.\label{SN10ae_optphot}}
\tablehead{
\colhead{JD$-2,455,000+$} &
\colhead{$g'$ (mag)}&
\colhead{$r'$ (mag)}&
\colhead{$i'$ (mag)}&
\colhead{$z'$ (mag)}&
\colhead{$B$ (mag)} &
\colhead{$V$ (mag)} &
\colhead{$Unfiltered$ (mag)}}
\startdata
249.6  &  $\cdots$      & $\cdots$     & $\cdots$     & $\cdots$       & $\cdots$     & $\cdots$    & 17.294(036)  \\
250.6  &  17.643(025) &  17.190(015)   & 17.114(027)  & 17.061(029)    & 17.799(034)  & 17.406(018) & 17.021(019)  \\   
251.5  &  17.522(022) &  17.054(016)   & 16.976(016)  & 16.849(025)    & 17.672(032)  & 17.292(022) & 16.841(023)  \\
252.5  &  17.446(020) &  $\cdots$      & $\cdots$     & $\cdots$       & 17.641(029)  & $\cdots$    & 16.802(015)  \\
253.6  &  $\cdots$    &  $\cdots$      & $\cdots$     & $\cdots$       & 17.682(037)  & 17.201(021) & $\cdots$     \\
254.6  &  17.529(033) &  16.928(019)   & 16.856(022)  & 16.639(020)    & 17.748(034)  & 17.094(035) & 16.732(015)  \\
255.7  &  $\cdots$    &  $\cdots$      & $\cdots$     & $\cdots$       & 17.883(085)  & 17.185(031) & $\cdots$     \\
256.6  &  17.691(042) &  16.942(016)   & 16.820(024)  & 16.610(018)    & 18.034(047)  & $\cdots$    & 16.825(016)  \\
258.7  &  $\cdots$    &  $\cdots$      & $\cdots$     & $\cdots$       & 18.685(049)  & 17.347(024) & $\cdots$     \\
259.5  &  18.139(029) &  $\cdots$      & $\cdots$     & $\cdots$       & 18.859(042)  & 17.446(019) & 17.019(023)  \\
260.5  &  $\cdots$    &  17.074(016)   & 16.883(015)  & 16.624(017)    & 19.214(059)  & 17.563(027) & 17.075(021)  \\
261.5  &  18.360(150) &  17.200(025)   & 16.926(025)  & 16.680(020)    & 19.248(063)  & 17.644(052) & 17.144(029)  \\
262.5  &  $\cdots$    &  17.253(017)   & 16.937(017)  & $\cdots$       & 19.488(062)  & 17.789(037) & 17.234(020)  \\
263.5  &  18.519(048) &  17.334(028)   & 17.050(067)  & 16.828(031)    & $\cdots$     & 17.624(032) & 17.338(030)  \\
264.5  &  18.742(039) &  $\cdots$      & $\cdots$     & $\cdots$       & $\cdots$     & $\cdots$    & 17.448(022)  \\
265.5  &  $\cdots$    &  $\cdots$      & 17.255(024)  & $\cdots$       & 19.851(114)  & $\cdots$    & 17.528(030)  \\
267.7  &  18.889(063) &  $\cdots$      & $\cdots$     & 17.035(026)    & $\cdots$     & 18.191(030) & $\cdots$     \\
269.5  &  $\cdots$    &  $\cdots$      & $\cdots$     & $\cdots$       & $\cdots$     & $\cdots$    & 17.887(029)  \\
270.7  &  $\cdots$    &  $\cdots$      & $\cdots$     & $\cdots$       & $\cdots$     & $\cdots$    & 18.110(109)  \\
271.5\tablenotemark{a}&  $\cdots$      & 17.966(033)  & 17.534(034)    & $\cdots$       & $\cdots$     & $\cdots$    & 18.116(056)  \\
273.6  &  $\cdots$    &  $\cdots$      & $\cdots$     & 17.387(035)    & $\cdots$     & 18.516(048) & $\cdots$     \\
274.5  &  $\cdots$    &  $\cdots$     & $\cdots$     & $\cdots$        & $\cdots$     & $\cdots$    & 18.224(057)  \\
276.6  &  19.483(085) &  $\cdots$      & $\cdots$     & $\cdots$       & 20.298(177)  & $\cdots$    & 18.384(072)  \\     
277.5  &  $\cdots$    &  $\cdots$      & $\cdots$     & $\cdots$       & $\cdots$     & $\cdots$    & 18.441(067)  \\
278.5  &  $\cdots$    &  $\cdots$      & $\cdots$     & $\cdots$       & $\cdots$     & $\cdots$    & 18.426(032)  \\
280.5  &  $\cdots$    &  $\cdots$      & $\cdots$     & $\cdots$       & $\cdots$     & $\cdots$    & 18.712(086)  \\
284.4  &  $\cdots$    &  $\cdots$      & $\cdots$     & $\cdots$       & $\cdots$     & $\cdots$    & 18.807(102)  \\
286.5  &  $\cdots$    &  $\cdots$      & $\cdots$     & $\cdots$       & $\cdots$     & $\cdots$    & 18.947(074)  \\
288.5  &  19.575(092) &  $\cdots$      & $\cdots$     & $\cdots$       & $\cdots$     & 19.176(072) & $\cdots$     \\
289.5  &  $\cdots$    &  18.868(062)   & 18.754(053)  & $\cdots$       & $\cdots$     & $\cdots$    & $\cdots$     \\    
291.5  &  $\cdots$    &  $\cdots$      & $\cdots$     & 17.988(039)    & 20.758(154)  & $\cdots$    & 19.320(154)  \\    
292.5  &  $\cdots$    &  $\cdots$      & $\cdots$     & $\cdots$       & $\cdots$     & $\cdots$    & $\cdots$     \\
296.5  &  $\cdots$    &  $\cdots$      & $\cdots$     & $\cdots$       & $\cdots$     & $\cdots$    & 19.556(195)  \\
297.5  &  19.881(157) &  19.070(054)   & 18.869(065)  & $\cdots$       & $\cdots$     & 19.671(133) & $\cdots$     \\
298.5  &  $\cdots$    &  $\cdots$      & $\cdots$     & $\cdots$       & 20.881(211)  & $\cdots$    & $\cdots$     \\    
299.5  &  $\cdots$    &  $\cdots$      & $\cdots$     & 18.397(079)    & $\cdots$     $\cdots$      & $\cdots$     \\
305.5\tablenotemark{a}& 20.174(089) &  19.293(049)   & $\cdots$     & $\cdots$       & $\cdots$     & $\cdots$    & $\cdots$     \\     
307.6  &  $\cdots$    &  $\cdots$      & $\cdots$     & $\cdots$       & $\cdots$     & $\cdots$    & 19.611(111)  \\
308.5  &  20.088(101) &  19.303(081)   & $\cdots$     & $\cdots$       & $\cdots$     & 19.678(153) & $\cdots$     \\
309.5  &  $\cdots$    &  $\cdots$      & 19.230(100)  & $\cdots$       & $\cdots$     & $\cdots$    & $\cdots$     \\    
313.5  &  $\cdots$    &  $\cdots$      & $\cdots$     & $\cdots$       & $\cdots$     & $\cdots$    & 19.863(190) \\
\enddata 
\tablecomments{Values in parentheses are 1$\sigma$ measurement uncertainties in millimag.}
\tablenotetext{a}{Photometry obtained from Swope images.}
\end{deluxetable}

%% file: SN10ae_nirphot.tex
  \begin{deluxetable}{ccccc}
\tablewidth{0pt}
\tabletypesize{\scriptsize}
\tablecaption{NIR photometry of SN~2010ae in the standard system.\label{SN10ae_nirphot}}
\tablehead{
\colhead{JD$-$$2,455,000+$} &
\colhead{$Y$ (mag)}&
\colhead{$J$ (mag)}&
\colhead{$H$ (mag)}& 
\colhead{Telescope\tablenotemark{a}}}
\startdata
258.6   &    16.207(015) &  16.128(015) &  15.717(015) & DUP \\ 
259.6   &    16.095(015) &  16.148(015) &  15.674(015) & DUP \\   
260.6   &    16.092(015) &  16.135(015) &  15.668(015) & DUP \\
261.6   &    16.066(015) &  16.171(015) &  15.661(015) & DUP \\
268.6   &    16.362(015) &  16.540(015) &  16.043(025) & SWO \\
269.6   &    16.400(019) &  16.596(015) &  16.078(022) & SWO \\
272.6   &    16.641(019) &  $\cdots$    &   $\cdots$   & SWO \\
273.6   &    16.627(015) &  16.837(016) &  16.186(032) & SWO \\
276.5   &    16.756(024) &  17.035(021) &  16.481(030) & SWO \\
\enddata 
\tablecomments{Values in parentheses are 1$\sigma$ measurement uncertainties in millimag.}
\tablenotetext{a}{DUP and SWO correspond to the du Pont and Swope telescopes, respectively.} 
\end{deluxetable}

%% file: specjournal.tex
\begin{deluxetable}{cccccclccc}
\tabletypesize{\scriptsize}
\tablewidth{0pt}
\tablenum{4}
\tablecaption{Journal of Spectroscopic Observations.\label{specjor}}
\tablehead{
\colhead{Date} &
\colhead{JD$-$2,455,000+} &
\colhead{Phase$^{a}$} &
\colhead{Telescope} &
\colhead{Instrument} &
\colhead{Grating} &
\colhead{Range} &
\colhead{Resolution} &
\colhead{No. of exposures} &
\colhead{Integration} \\ 
\colhead{} &
\colhead{} &
\colhead{} &
\colhead{} &
\colhead{} &
\colhead{} &
\colhead{(\AA)} &
\colhead{(FWHM \AA)} &
\colhead{} &
\colhead{(s)}}
\startdata
2010 Feb 23  & 250.57 & $-$2.1   & GEM-S  & GMOS      & R600         & 4858--7729  & 8      &  1        & 1200\\
2010 Feb 24  & 251.51 & $-$1.1   & VLT    & XSHOOTER  &UV$+$VIS$+$NIR& 3150--24790 &        &  8, 8, 16 & 200, 200, 100 \\
2010 Feb 25  & 252.52 & $-$0.1   & GEM-S  & GMOS      & B600$+$R600  & 3583--8959  & 8      &  1, 1     & 600 \\
2010 Feb 26  & 253.52 & $+$0.4   & GEM-S  & GMOS      & B600$+$R600  & 3583--8959  & 8      &  1, 1     & 600 \\
2010 Feb 27  & 254.52 & $+$1.9   & GEM-S  & GMOS      & B600$+$R600  & 3581--8960  & 8      &  1, 1     & 800 \\
2010 Feb 28  & 255.51 & $+$2.9   & GEM-S  & GMOS      & B600$+$R600  & 3583--8959  & 8      &  1, 1     & 800 \\
2010 Mar 01  & 256.51 & $+$3.9   & GEM-S  & GMOS      & B600$+$R600  & 3583--8959  & 8      &  1, 1     & 800 \\
2010 Mar 02  & 257.67 & $+$5.0   & GEM-S  & GMOS      & B600$+$R600  & 3583--8959  & 8      &  1, 1     & 850 \\
2010 Mar 04  & 259.57 & $+$6.9   & GEM-S  & GMOS      & B600$+$R600  & 3582--8959  & 8      &  1, 1     & 700 \\
2010 Mar 05  & 260.52 & $+$7.9   & NTT    & EFOSC     &gm\#11+gm\#16 & 3352--10291 & 14     &  1, 1     & 600\\ 
2010 Mar 06  & 261.55 & $+$8.9   & VLT    & XSHOOTER  &UV$+$VIS$+$NIR& 3150--24790 & 0.8, 0.8, 3.2&  4, 4, 8  & 600, 600, 300\\ 
2010 Mar 07  & 262.51 & $+$9.9   & VLT    & XSHOOTER  &UV$+$VIS$+$NIR& 3150--24790 & 0.8, 0.8, 3.2&  4, 4, 8  & 600, 600, 300\\ 
2010 Mar 08  & 263.52 & $+$10.9  & NTT    & SOFI      &GB$+$GR       & 9358--25000 & 27, 30 &  12, 20  & 120 \\ 
2010 Mar 08  & 263.55 & $+$10.9  & VLT    & XSHOOTER  &UV$+$VIS$+$NIR& 3150--24790 & 0.8, 0.8, 3.2&  4, 4, 8  & 600, 600, 300\\                       
2010 Mar 10  & 265.53 & $+$12.9  & VLT    & XSHOOTER  &UV$+$VIS$+$NIR& 3150--24790 & 0.8, 0.8, 3.2&  4, 4, 8  & 600, 600, 300\\                       
2010 Mar 12  & 267.55 & $+$14.9  & SOAR   & GOODMAN   &RALC300       & 3556--8884  & 0.8, 0.8, 3.2&  3 & 600 \\
2010 Mar 13  & 268.55 & $+$15.9  & VLT    & XSHOOTER  &UV$+$VIS$+$NIR& 3150--24790 & 0.8, 0.8, 3.2&  4, 4, 8  & 600, 600, 300\\                      
2010 Mar 15  & 270.54 & $+$17.9  & VLT    & XSHOOTER  &UV$+$VIS$+$NIR& 3150--24790 & 0.8, 0.8, 3.2&  4, 4, 8  & 600, 600, 300\\
2010 Mar 15  & 270.74 & $+$18.1  & DUP    & WFCCD     & B            & 3674--9254  & 7      &  3 & 600 \\
2010 Mar 20  & 275.66 & $+$23.0  & DUP    & WFCCD     & B            & 3674--9252  & 7      &  1 & 900 \\
2010 Apr 17  & 304.50 & $+$51.9  & GEM-S  & GMOS      & B600         & 3582--6420  & 8      &  1 & 2500\\
2010 Apr 22  & 309.50 & $+$56.9  & GEM-S  & GMOS      & R400         & 5378--9636  & 8      &  1 & 2500\\
2010 Nov 04  & 504.81 & $+$252.2 & VLT    & FORS2     & 300V         & 3150--8883  & 9    &  1 & 1793\\
\enddata
\tablenotetext{a}{Days since $B_{max}$ (JD$-$2455252.65).}
\end{deluxetable}

%% file: lcparameters.tex
\begin{deluxetable} {lcclc}
\tablecolumns{5}
\tablenum{5}
\tablewidth{0pc}
\tablecaption{Lightcurve Parameters of SNe~2008ha and 2010ae.\label{lcpar}}
\tablehead{
\colhead{Filter} &
\colhead{Peak Time} &
\colhead{Peak Obs.} &
\colhead{Peak Abs.} &
\colhead{$\Delta$$m_{15}$} \\
\colhead{} &
\colhead{(JD$-$2,450,000)} &
\colhead{(mag)} &
\colhead{(mag)} &
\colhead{(mag)} }
\startdata
\multicolumn{5}{c}{\bf SN 2008ha}\\
$u'$     & 4782.23$\pm$1.40 & 19.33$\pm$0.12  & $-$12.64$\pm$0.16 & 2.74$\pm$0.31\\
$B$      & 4782.43$\pm$1.50 & 18.13$\pm$0.06  & $-$13.79$\pm$0.14 & 2.03$\pm$0.20  \\
$g'$     & 4783.76$\pm$0.22&  17.87$\pm$0.02  & $-$14.01$\pm$0.14 & 1.80$\pm$0.03 \\
$V$      & 4785.67$\pm$0.48&  17.74$\pm$0.01  & $-$14.11$\pm$0.14 & 1.29$\pm$0.04 \\
$r'$     & 4788.15$\pm$0.51&  17.67$\pm$0.01  & $-$15.15$\pm$0.14 & 1.11$\pm$0.04 \\
$i'$     & 4789.05$\pm$0.55&  17.74$\pm$0.02  & $-$14.04$\pm$0.14 & 0.85$\pm$0.04 \\
$Y$      & 4790.31$\pm$0.79&  17.50$\pm$0.04  & $-$14.21$\pm$0.14 & $\cdots$      \\
$J$      & 4791.01$\pm$1.27&  17.72$\pm$0.04  & $-$13.98$\pm$0.14 & $\cdots$      \\
$H$      & 4792.01$\pm$2.41&  17.18$\pm$0.08  & $-$14.50$\pm$0.15 & $\cdots$      \\
\multicolumn{5}{c}{\bf SN 2010ae}\\
$B$      & 5252.65$\pm$0.20 & 17.65$\pm$0.02  & $-13.44 \gtrsim M_{B} \gtrsim$ $-$15.47$\pm$0.54 & 2.43$\pm$0.11 \\
$g'$     & 5253.00$\pm$0.18 & 17.49$\pm$0.02  & $-13.54 \gtrsim M_{g} \gtrsim$ $-$15.33$\pm$0.54 & 1.51$\pm$0.05 \\
$V$      & 5254.74$\pm$0.18 & 17.17$\pm$0.02  & $-13.80 \gtrsim M_{V} \gtrsim$ $-$15.33$\pm$0.54 & 1.15$\pm$0.04 \\
$r'$     & 5255.78$\pm$0.30 & 16.92$\pm$0.01  & $-13.99 \gtrsim M_{r} \gtrsim$ $-$15.29$\pm$0.54 & 1.01$\pm$0.03\\
$i'$     & 5257.04$\pm$0.23 & 16.82$\pm$0.01  & $-14.01 \gtrsim M_{i} \gtrsim$ $-$14.99$\pm$0.54 & 0.80$\pm$0.04\\
$z'$     & 5257.70$\pm$0.31 & 16.58$\pm$0.02  & $-14.18 \gtrsim M_{z} \gtrsim$ $-$14.88$\pm$0.54 & 0.76$\pm$0.04 \\
$Y$      & 5261.47$\pm$0.72 & 16.07$\pm$0.02  & $-14.64 \gtrsim M_{Y} \gtrsim$ $-$15.16$\pm$0.54 & $\cdots$      \\
$J$      & 5258.75$\pm$0.53 & 16.13$\pm$0.05  & $-14.56 \gtrsim M_{J} \gtrsim$ $-$14.99$\pm$0.54 & 0.74$\pm$0.03 \\
$H$      & 5260.97$\pm$0.80 & 15.66$\pm$0.02  & $-14.99 \gtrsim M_{H} \gtrsim$ $-$15.26$\pm$0.54 & 0.77$\pm$0.07 \\
\enddata
\end{deluxetable}

%% file: SN08ha_locseq.tex
\begin{deluxetable} {cccccccccccc}
\rotate
\tabletypesize{\tiny}
\tablecolumns{12}
\tablewidth{0pt}
\tablecaption{Optical and NIR photometry of the local sequence of SN~2008ha in the standard system.\label{SN08halocalsequence}}
\tablehead{
\colhead{}     &
\colhead{}     &
\colhead{}     &
\colhead{$u'$} &
\colhead{$g'$} &
\colhead{$r'$} &
\colhead{$i'$} &
\colhead{$B$}  &
\colhead{$V$}  &
\colhead{$Y$}  &
\colhead{$J$}  &
\colhead{$H$} \\
\colhead{STAR}             &
\colhead{$\alpha~(2000)$}  &
\colhead{$\delta~(2000)$}  &
\colhead{(mag)}            &
\colhead{(mag)}            &
\colhead{(mag)}            &
\colhead{(mag)}            &
\colhead{(mag)}            &
\colhead{(mag)}            &
\colhead{(mag)}            &
\colhead{(mag)}            &
\colhead{(mag)}}
\startdata
01 & 23$^{\rm h}$34$^{\rm m}$35$\fs$95 & $+$18$^{\circ}$12$\arcmin$38$\farcs$84  &  18.929(026)&  16.912(011)	&  16.230(011)	&  16.017(011)	&  17.399(018)	&  16.515(010)	& $ \cdots $	& $ \cdots $	& $ \cdots $	\\
02 & 23$^{\rm h}$35$^{\rm m}$03$\fs$38 & $+$18$^{\circ}$16$\arcmin$33$\farcs$89  &  17.072(040)&  15.163(012)	&  14.431(018)	&  14.166(011)	&  15.672(020)	&  14.732(010)	&  13.441(104)	&  13.133(045)	&  12.711(069)	\\
03 & 23$^{\rm h}$34$^{\rm m}$53$\fs$76 & $+$18$^{\circ}$10$\arcmin$46$\farcs$81  &  15.778(021)& $ \cdots $	& $ \cdots $	& $ \cdots $	&  14.734(008)	& $ \cdots $	&  13.042(009)	&  12.843(012)	&  12.522(011)	\\
04 & 23$^{\rm h}$34$^{\rm m}$47$\fs$41 & $+$18$^{\circ}$17$\arcmin$39$\farcs$86  &  17.098(071)& $ \cdots $	& $ \cdots $	& $ \cdots $	&  15.147(016)	& $ \cdots $	&  12.195(008)	&  11.884(016)	&  11.307(020)	\\
05 & 23$^{\rm h}$34$^{\rm m}$43$\fs$55 & $+$18$^{\circ}$12$\arcmin$38$\farcs$84  &  19.099(082)&  16.173(009)	&  14.832(008)	&  14.189(010)	&  16.876(008)	&  15.457(019)	&  13.060(010)	&  12.688(014)	&  12.059(008)	\\
06 & 23$^{\rm h}$34$^{\rm m}$37$\fs$55 & $+$18$^{\circ}$10$\arcmin$31$\farcs$82  &  17.076(024)&  15.633(007)	&  15.071(007)	&  14.861(008)	&  16.047(008)	&  15.294(007)	& $ \cdots $	& $ \cdots $	& $ \cdots $	\\
07 & 23$^{\rm h}$34$^{\rm m}$55$\fs$68 & $+$18$^{\circ}$16$\arcmin$25$\farcs$85  &  18.030(030)&  15.873(007)	&  15.114(007)	&  14.857(008)	&  16.388(008)	&  15.443(009)	&  14.050(009)	&  13.800(010)	&  13.355(017)	\\
08 & 23$^{\rm h}$34$^{\rm m}$43$\fs$29 & $+$18$^{\circ}$14$\arcmin$52$\farcs$89  &  17.034(018)&  15.788(007)	&  15.277(007)	&  15.068(008)	&  16.174(008)	&  15.484(007)	&  14.326(010)	&  14.110(023)	&  13.776(025)	\\
09 & 23$^{\rm h}$34$^{\rm m}$45$\fs$84 & $+$18$^{\circ}$10$\arcmin$45$\farcs$86  &  18.854(033)&  16.378(007)	&  15.431(007)	&  15.083(008)	&  16.952(008)	&  15.869(007)	&  14.182(009)	&  13.845(010)	&  13.275(010)	\\
10 & 23$^{\rm h}$34$^{\rm m}$58$\fs$98 & $+$18$^{\circ}$15$\arcmin$59$\farcs$85  &  20.017(054)&  16.912(008)	&  15.513(009)	&  14.781(009)	&  17.661(011)	&  16.160(014)	&  13.605(008)	&  13.222(015)	&  12.559(020)	\\
11 & 23$^{\rm h}$35$^{\rm m}$01$\fs$21 & $+$18$^{\circ}$10$\arcmin$10$\farcs$80  &  18.653(051)&  16.402(007)	&  15.624(007)	&  15.345(008)	&  16.936(008)	&  15.954(007)	&  14.505(010)	&  14.203(018)	&  13.738(017)	\\
12 & 23$^{\rm h}$34$^{\rm m}$36$\fs$90 & $+$18$^{\circ}$17$\arcmin$07$\farcs$84  &  18.444(031)&  16.576(007)	&  15.842(007)	&  15.555(008)	&  17.067(008)	&  16.159(009)	& $ \cdots $	& $ \cdots $	& $ \cdots $	\\
13 & 23$^{\rm h}$34$^{\rm m}$52$\fs$61 & $+$18$^{\circ}$13$\arcmin$22$\farcs$83  &  17.971(020)&  16.666(008)	&  16.179(007)	&  16.003(008)	&  17.053(008)	&  16.378(007)	& $ \cdots $	& $ \cdots $	& $ \cdots $	\\
14 & 23$^{\rm h}$34$^{\rm m}$41$\fs$30 & $+$18$^{\circ}$10$\arcmin$57$\farcs$85  &  17.974(016)&  16.851(021)	&  16.402(008)	&  16.230(008)	&  17.203(012)	&  16.586(010)	& $ \cdots $	& $ \cdots $	& $ \cdots $	\\
15 & 23$^{\rm h}$34$^{\rm m}$55$\fs$05 & $+$18$^{\circ}$17$\arcmin$04$\farcs$89  &  18.398(030)&  16.967(008)	&  16.431(007)	&  16.241(008)	&  17.367(011)	&  16.643(012)	&  15.530(008)	&  15.307(014)	&  14.967(011)	\\
16 & 23$^{\rm h}$34$^{\rm m}$36$\fs$86 & $+$18$^{\circ}$11$\arcmin$23$\farcs$86  &  20.478(066)&  17.672(009)	&  16.646(007)	&  16.280(008)	&  18.276(012)	&  17.117(007)	& $ \cdots $	& $ \cdots $	& $ \cdots $	\\
17 & 23$^{\rm h}$34$^{\rm m}$47$\fs$99 & $+$18$^{\circ}$11$\arcmin$00$\farcs$86  &  19.158(049)&  17.430(009)	&  16.774(007)	&  16.536(008)	&  17.888(010)	&  17.047(009)	&  15.757(010)	&  15.457(019)	&  15.053(018)	\\
18 & 23$^{\rm h}$34$^{\rm m}$52$\fs$89 & $+$18$^{\circ}$11$\arcmin$09$\farcs$89  &  18.686(027)&  17.418(008)	&  16.962(007)	&  16.805(008)	&  17.782(009)	&  17.139(007)	&  16.134(008)	&  15.878(015)	&  15.556(017)	\\
19 & 23$^{\rm h}$34$^{\rm m}$45$\fs$73 & $+$18$^{\circ}$16$\arcmin$17$\farcs$88  &  18.974(027)&  17.525(008)	&  16.955(010)	&  16.742(008)	&  17.948(012)	&  17.196(011)	& $ \cdots $	& $ \cdots $	& $ \cdots $	\\
20 & 23$^{\rm h}$35$^{\rm m}$07$\fs$11 & $+$18$^{\circ}$17$\arcmin$56$\farcs$84  &  18.749(057)&  17.498(030)	&  16.945(014)	&  16.760(016)	&  17.855(026)	&  17.144(014)	& $ \cdots $	& $ \cdots $	& $ \cdots $	\\
21 & 23$^{\rm h}$34$^{\rm m}$38$\fs$28 & $+$18$^{\circ}$12$\arcmin$28$\farcs$87  &  19.757(119)&  17.834(009)	&  17.082(007)	&  16.781(008)	&  18.363(012)	&  17.384(019)	& $ \cdots $	& $ \cdots $	& $ \cdots $	\\
22 & 23$^{\rm h}$34$^{\rm m}$59$\fs$72 & $+$18$^{\circ}$16$\arcmin$53$\farcs$80  &  19.652(055)&  18.409(010)	&  17.071(007)	&  16.483(014)	&  19.103(038)	&  17.714(010)	&  15.386(012)	&  14.988(012)	&  14.373(009)	\\
23 & 23$^{\rm h}$34$^{\rm m}$49$\fs$31 & $+$18$^{\circ}$14$\arcmin$23$\farcs$80  & $ \cdots $	&  18.270(009)	&  17.214(007)	&  16.832(009)	&  18.902(016)	&  17.713(012)	&  15.916(007)	&  15.552(018)	&  14.961(012)	\\
24 & 23$^{\rm h}$34$^{\rm m}$59$\fs$47 & $+$18$^{\circ}$13$\arcmin$27$\farcs$88  & $ \cdots $	&  18.628(009)	&  17.265(010)	&  15.957(016)	&  19.464(026)	&  17.862(016)	&  14.441(007)	&  14.007(009)	&  13.408(009)	\\
25 & 23$^{\rm h}$35$^{\rm m}$08$\fs$38 & $+$18$^{\circ}$15$\arcmin$26$\farcs$88  & $ \cdots $	&  18.756(009)	&  17.400(014)	&  16.433(015)	&  19.569(027)	&  18.034(026)	&  15.153(016)	&  14.735(014)	&  14.106(008)	\\
26 & 23$^{\rm h}$34$^{\rm m}$39$\fs$18 & $+$18$^{\circ}$14$\arcmin$36$\farcs$85  & $ \cdots $	&  18.827(009)	&  17.836(008)	&  17.455(013)	&  19.386(024)	&  18.283(017)	& $ \cdots $	& $ \cdots $	& $ \cdots $	\\
27 & 23$^{\rm h}$35$^{\rm m}$08$\fs$00 & $+$18$^{\circ}$12$\arcmin$44$\farcs$81  & $ \cdots $	&  18.882(009)	&  17.772(008)	&  17.307(009)	&  19.505(027)	&  18.278(035)	& $ \cdots $	& $ \cdots $	& $ \cdots $	\\
28 & 23$^{\rm h}$34$^{\rm m}$55$\fs$17 & $+$18$^{\circ}$18$\arcmin$43$\farcs$84  & $ \cdots $	& $ \cdots $	& $ \cdots $	& $ \cdots $	& $ \cdots $	& $ \cdots $	&  13.175(008)	&  12.935(025)	&  12.554(015)	\\
29 & 23$^{\rm h}$35$^{\rm m}$16$\fs$19 & $+$18$^{\circ}$18$\arcmin$00$\farcs$86  & $ \cdots $	& $ \cdots $	& $ \cdots $	& $ \cdots $	& $ \cdots $	& $ \cdots $	&  14.059(010)	&  13.806(021)	&  13.405(010)	\\
30 & 23$^{\rm h}$35$^{\rm m}$13$\fs$81 & $+$18$^{\circ}$11$\arcmin$39$\farcs$82  & $ \cdots $	& $ \cdots $	& $ \cdots $	& $ \cdots $	& $ \cdots $	& $ \cdots $	&  14.337(007)	&  14.035(015)	&  13.564(011)	\\
31 & 23$^{\rm h}$34$^{\rm m}$49$\fs$67 & $+$18$^{\circ}$18$\arcmin$54$\farcs$81  & $ \cdots $	& $ \cdots $	& $ \cdots $	& $ \cdots $	& $ \cdots $	& $ \cdots $	&  14.303(009)	&  14.055(019)	&  13.665(021)	\\
32 & 23$^{\rm h}$34$^{\rm m}$43$\fs$52 & $+$18$^{\circ}$19$\arcmin$09$\farcs$90  & $ \cdots $	& $ \cdots $	& $ \cdots $	& $ \cdots $	& $ \cdots $	& $ \cdots $	&  14.994(017)	&  14.772(034)	&  14.447(012)	\\
33 & 23$^{\rm h}$35$^{\rm m}$13$\fs$13 & $+$18$^{\circ}$12$\arcmin$05$\farcs$88  & $ \cdots $	& $ \cdots $	& $ \cdots $	& $ \cdots $	& $ \cdots $	& $ \cdots $	&  15.717(011)	&  15.211(012)	&  14.536(014)	\\
34 & 23$^{\rm h}$35$^{\rm m}$01$\fs$45 & $+$18$^{\circ}$09$\arcmin$36$\farcs$89  & $ \cdots $	& $ \cdots $	& $ \cdots $	& $ \cdots $	& $ \cdots $	& $ \cdots $	&  15.882(012)	&  15.367(028)	&  14.866(027)	\\
35 & 23$^{\rm h}$34$^{\rm m}$44$\fs$67 & $+$18$^{\circ}$08$\arcmin$57$\farcs$81  & $ \cdots $	& $ \cdots $	& $ \cdots $	& $ \cdots $	& $ \cdots $	& $ \cdots $	&  15.908(014)	&  15.423(032)	&  14.914(048)	\\
36 & 23$^{\rm h}$35$^{\rm m}$09$\fs$00 & $+$18$^{\circ}$18$\arcmin$03$\farcs$84  & $ \cdots $	& $ \cdots $	& $ \cdots $	& $ \cdots $	& $ \cdots $	& $ \cdots $	&  16.116(016)	&  15.714(024)	&  15.096(012)	\\
37 & 23$^{\rm h}$35$^{\rm m}$12$\fs$94 & $+$18$^{\circ}$19$\arcmin$15$\farcs$89  & $ \cdots $	& $ \cdots $	& $ \cdots $	& $ \cdots $	& $ \cdots $	& $ \cdots $	&  16.304(012)	&  15.811(020)	&  15.323(023)	\\
\enddata
\tablecomments{Uncertainties given in parentheses in thousandths of a
  magnitude correspond to an rms of the magnitudes obtained on
  photometric nights.}
\end{deluxetable}

%% file: SN08ha_optphot.tex
\clearpage
\begin{deluxetable}{cccccccc}
\tablewidth{0pt}
\tabletypesize{\scriptsize}
\tablecaption{Swope optical photometry of SN~2008ha in the standard system.\label{SN08ha_optphot}}
\tablehead{
\colhead{JD$-2,454,000+$} &
\colhead{$u'$ (mag)}&
\colhead{$g'$ (mag)}&
\colhead{$r'$ (mag)}&
\colhead{$i'$ (mag)}&
\colhead{$B$ (mag)} &
\colhead{$V$ (mag)} }
\startdata
781.6 &   19.332(138) &  17.992(041) &  18.004(044) &  18.104(044) &  18.144(049) &  17.911(035)\\ 
783.6 &   19.390(131) &  17.853(032) &  17.868(028) &  17.963(033) &  18.154(054) &  17.771(036)\\
785.6 &   19.644(098) &  17.936(018) &  17.701(019) &  17.799(023) &  18.294(039) &  17.746(020)\\
786.5 &   19.922(061) &  18.027(015) &  17.709(015) &  17.800(015) &  18.393(020) &  17.738(016)\\
789.6 &   20.574(080) &  18.352(016) &  17.692(020) &  17.730(028) &  18.878(030) &  17.918(020)\\
792.5 &   21.181(125) &  18.717(106) &  17.861(031) &  17.814(044) &  19.447(045) &  18.208(028)\\
793.5 &   21.235(146) &  19.048(144) &  18.009(075) &  17.950(042) &  19.612(061) &  18.344(027)\\
795.5 &   21.623(188) &  19.247(127) &  18.242(035) &  18.136(034) &  19.977(073) &  18.600(041)\\
797.5 &   22.207(335) &  19.571(029) &  18.425(022) &  18.243(025) &  20.244(203) &  18.849(168)\\
798.5 &   22.482(395) &  19.637(029) &  18.476(023) &  18.317(033) &  20.147(145) &  18.848(096)\\
801.6 &   22.591(357) &  19.822(035) &  18.694(028) &  18.476(036) &  20.490(085) &  19.043(046)\\
802.5 &   $\cdots$    &  19.935(266) &  18.738(064) &  18.521(056) &  20.538(071) &  19.136(034)\\
803.5 &   $\cdots$    &  19.907(037) &  18.813(031) &  18.565(034) &  20.722(178) &  19.118(085)\\
806.5 &   $\cdots$    &  20.308(262) &  18.952(058) &  18.695(064) &  $\cdots$    &  19.301(064)\\
807.5 &   $\cdots$    &  20.174(357) &  19.092(113) &  18.715(083) &  $\cdots$    &  19.286(091)\\
808.5 &   $\cdots$    &  $\cdots$    &  19.150(108) &  18.987(120) &  $\cdots$    &  19.498(136)\\
809.5 &   $\cdots$    &  20.491(387) &  19.272(075) &  18.915(095) &  $\cdots$    &  19.593(116)\\
810.5 &   $\cdots$    &  $\cdots$    &  19.250(080) &  18.956(087) &  $\cdots$    &  19.509(096)\\
\enddata 
\tablecomments{Values in parentheses are 1$\sigma$ measurement uncertainties in millimag.}
\end{deluxetable}

%% file: SN08ha_nirphot.tex
  \begin{deluxetable}{ccccc}
\tablewidth{0pt}
\tabletypesize{\scriptsize}
\tablecaption{Swope NIR photometry of SN~2008ha in the standard system.\label{SN08ha_nirphot}}
\tablehead{
\colhead{JD$-$$2,454,000+$} &
\colhead{$Y$ (mag)}&
\colhead{$J$ (mag)}&
\colhead{$H$ (mag)}}
\startdata
782.6 &  17.832(096) &17.819(438) & 17.652(115) \\ 
784.6 &  17.872(090) &17.768(058) & 17.468(130) \\ 
787.6 &  17.540(021) &17.663(046) & 17.281(072) \\ 
791.5 &  17.504(030) &17.674(062) & $\cdots$    \\ 
800.5 &  17.945(045) &17.824(116) & 17.552(121) \\ 
\enddata 
\tablecomments{Values in parentheses are 1$\sigma$ measurement uncertainties in millimag.} 
\end{deluxetable}